\documentstyle[12pt, amsmath, amsfonts, epsfig, amssymb, rotating]{article}
\begin{document}
\def\be{\begin{equation}} \def\ovx{\vec{x}} \def\ovy{\vec{y}}
\def\ovr{\vec{r}} \def\ovn{\vec{n}} \def\vl{\vec{L}} \def\vp{\varphi}
\def\eps{\epsilon} \def\kt{K_{2}(\tau)} \def\cala{{\mathcal A}}
\def\caln{{\mathcal N}} \def\cald{{\mathcal D}} \def\calc{{\mathcal C}}
\def\calr{{\mathcal R}} \def\calh{{\mathcal H}} \def\calf{{\mathcal F}}
\def\pd{\pi/2 } \def\fpd{\frac{\pi}{2}} \def\pt{\frac{\pi}{3}}
\def\pn{\frac{\pi}{n}} \def\rm{\mathbb{R}} \def\nm{\mathbb{N}}
\def\cm{\mathbb{C}} \def\zm{\mathbb{Z}} \def\hm{\mathbb{H}} \def\im{{\mathcal
I}m\ } \def\dm{\frac{1}{2}} \def\ba{\bar{\alpha}} \def\equi{\displaystyle{\ \
\mathop{\sim}_{s\to 1}\ \ }} \def\dn{\cald^{\nu}} \def\pli{l_{1}\ldots
l_{\nu}} \def\pri{R_{1}\ldots R_{\nu}} \def\sli{l_1+\cdots+l_{\nu}}
\def\sri{R_{1}+\cdots+ R_{\nu}} \def\pliu{l_{1}\ldots l_{\nu_1}}
\def\priu{R_{1}\ldots R_{\nu_1}} \def\sliu{l_1+\cdots+l_{\nu_1}}
\def\sriu{R_{1}+\cdots+ R_{\nu_1}} \def\plid{l'_{1}\ldots l'_{\nu_2}}
\def\prid{R'_{1}\ldots R'_{\nu_2}} \def\slid{l'_1+\cdots+l'_{\nu_2}}
\def\srid{R'_{1}+\cdots+ R'_{\nu_2}} \def\plpi{l'_{1}\ldots l'_{\nu}}
\def\prpi{R'_{1}\ldots R'_{\nu}} \def\slpi{l'_1+\cdots+l'_{\nu}}
\def\srpi{R'_{1}+\cdots+ R'_{\nu}} \def\plinr{l_{1}\ldots l_{m}}
\def\prinr{R^{(r)}_{1}\ldots R^{(r)}_{m_r}} \def\slinr{l_1+\cdots+l_{m}}
\def\srinr{R_{1}+\cdots+ R_{m}} \def\plpinr{l'_{1}\ldots l'_{n}}
\def\prpinr{{R'}^{(r)}_{1}\ldots {R'}^{(r)}_{n_r}}
\def\slpinr{l'_1+\cdots+l'_{n}} \def\srpinr{R'_{1}+\cdots+ R'_{n_r}}
\def\sti{\tau_{1}+\cdots+ \tau_{\nu}} \def\ln{l_{\nu}} \def\tr{{\textrm tr}}
\def\cc{{\textrm c.c.}} \def\vdv{{}^{t}\bar{V} D V} \def\dm{\frac{1}{2}}

\title{\bf Semi-classical calculations of the two-point correlation form factor
  for diffractive systems}
\author{E. Bogomolny and O. Giraud\\
 Laboratoire de Physique Th\'eorique et Mod\`eles Statistiques 
 \thanks{Unit\'e Mixte de Recherche de l'Universit\'e Paris XI et du CNRS
   (UMR 8626)}\\
Universit\'e de Paris XI, B\^at. 100\\
91405 Orsay Cedex, France}

\maketitle

\begin{abstract}

The computation of the two-point correlation form factor $K(\tau)$ is
performed for a rectangular billiard with a small size impurity inside for
both  periodic or Dirichlet boundary conditions. It is demonstrated that all
terms of perturbation expansion of this form factor in powers of $\tau$ can
be computed directly from a semiclassical trace formula. The main part of the
calculation is the summation of non-diagonal terms in the cross product of
classical orbits. When the diffraction coefficient for the scattering on the
impurity is a constant our results coincide with the expansion of exact
expressions  obtained in 
Ref.~\cite{gerland} by a different method.
\end{abstract}

\pagebreak

\section{Introduction}
It is widely accepted that statistical properties of the energy levels of a
quantum problem are determined mainly by the nature of its classical motion.
Thus spectral statistics of classically chaotic systems follow the random
matrix predictions \cite{bohigasgiannoni}, \cite{mehta} and spectral statistics 
of classically integrable
models agree with the Poisson statistics \cite{berrytaborc}. A lot of 
numerical evidence (see e.g. \cite{bohigas}) and partial analytical arguments 
\cite{andreev}-\cite{marklof} strongly support these conjectures.

But there are models which (for various reasons) do not fit into the
above scheme. A noticeable example is given by the so-called diffractive
systems where one or a few small-size scatters are added to an otherwise
smooth model. In classical mechanics point-like singularities affect only 
a zero-measure set
of trajectories which directly hit them but in quantum mechanics singularities
often lead to important effects. Though the addition of a scatter to a chaotic
system has no effect on its spectral statistics \cite{leboeuf} even one
small-size impurity changes completely the statistical properties of integrable
models \cite{seba}.

The difference between chaotic and integrable models can intuitively be
understood as follows. In chaotic systems classical trajectories and quantum
wave functions cover the whole surface of constant energy and the addition
of a few centers of scattering can not change this property. On the other
hand, in integrable models trajectories and wave functions are confined to
invariant tori and small-size scatters will, in general, induce transitions
between different tori, qualitatively changing the nature of motion.

The investigation of integrable models with short-range scatters has been
initiated in Ref.~\cite{seba} (see also \cite{shigehara} and references
therein) and recently attracts wide attention from both
experimental and theoretical points of view. Experimentally,  non-hydrogenic 
atoms in external fields \cite{delande} are just a  possible realization of 
such models.
Theoretically, these models are rare examples where spectral statistics can
be computed analytically.

In Ref.~\cite{gerland} the two-point correlation function for a rectangular
billiard with a delta-function potential has been computed exactly and in
\cite{giraud} the nearest-neighbor distributions for this model were obtained.
The main conclusion of these works is that the spectral statistics of
diffractive systems belong to the so-called intermediate type. Intermediate 
spectral statistics describes
level repulsion like random matrix ensembles but the nearest-neighbor
distribution has exponential decrease like Poisson statistics.
These mixed properties of diffractive systems are similar to spectral
statistics of the 3-dimensional Anderson model at the metal-insulator
transition point \cite{shklovskii}. 

The purpose of this paper is the investigation of spectral statistics of
diffractive systems by semiclassical methods. Our starting point is the
trace formula for diffractive systems derived in \cite{rosenquist} and
\cite{pavloff} which allows to express the Green function and all related
quantities as a sum over periodic and diffractive orbits.
The two-point correlation function is obtained as usual by smoothing
the product of 2 densities of states over a small energy interval. It has
been shown in \cite{berry} that  the linear on $\tau$ term in the expansion
of  the two-point correlation form factor $K(\tau)$ can be
obtained by a diagonal approximation, which consists of taking into account
only contributions from pairs of orbits with exactly the same length, 
and summing them  using the Hannay-Ozorio de Almeida sum rule \cite{ozorio}. 

The main difficulty of obtaining the higher order terms of the expansion of
$K(\tau)$ in powers of $\tau$ is the necessity to take into account
non-diagonal contributions of classical orbits with slightly different
actions. The method which permits to solve this problem for integrable
systems has been proposed in \cite{bogomolny2}. Using and simplifying it we
construct all terms of perturbation series expansion of the two-point
correlation form factor $K(\tau)$ into series of $\tau$
for rectangular billiard with a delta-function impurity for both periodic and
Dirichlet boundary conditions. The results agree with the expansion of exact
formulas obtained in \cite{gerland}.

The model with a delta-function potential is characterized by a constant
diffraction coefficient. It is of interest to consider more general models 
with a non-constant
diffraction coefficient. Though the exact solution of such models are not
known we construct the perturbation series expansion of $K(\tau)$ in this
case as well. 

The plan of the paper is the following. The trace formula for diffractive
systems is shortly discussed in  Section \ref{section1}. In Section
\ref{section2} the detailed discussion of perturbation series for the form
factor of a rectangular billiard with delta-function potential and
Dirichlet boundary conditions is presented. The main part of this section is
devoted to the summation of non-diagonal contributions. In Section
\ref{section3} the case of Dirichlet boundary conditions is discussed and in
Section \ref{section4} the case of arbitrary diffraction coefficient is
considered. The expansion of the exact expression of $K(\tau)$ for
rectangular billiards with constant diffraction coefficient obtained in
\cite{gerland} is derived in Appendix.

\section{Trace formula for diffractive systems}\label{section1}

The starting point of the modern semiclassical approximation for
multi-di\-men\-sio\-nal
quantum systems is the semiclassical approximation for the (advanced)  Green
function as a sum over classical trajectories with energy $E$ starting from
initial point $\vec{x}$ with momentum in the direction $\vec{n}$ and ending
at final point $\vec{y}$ with the momentum in the direction $\vec{n}'$ 
\cite{bb}, \cite{Gutz}
\begin{equation}
\label{sumtraj}
G(\vec{x},\vec{y})=\sum_{tr} G((\vec{x},\vec{n}), (\vec{y},\vec{n}')),
\end{equation}
where the contribution from each trajectory has the form
\begin{equation}
 G((\vec{x},\vec{n}), (\vec{y},\vec{n}'))=A_{tr}
\exp (\frac{i}{\hbar}S_{cl}-i\frac{\pi}{2}\nu).
\end{equation}
$S_{cl}=S_{cl}(E, \vec{x},\vec{y})$ is  the classical action computed along 
the trajectory,
\be
A_{tr}=\frac{m}{i\hbar(2\pi \hbar)^{\dm(f-1)}}
  \left|\frac{1}{k k'}\det \frac{\partial^2 S_{cl}}{\partial y \partial y'}
\right|^{\dm}
\end{equation}
where $m$ is the mass, $f$ is the dimension of the space, $y$ and $ y'$ are 
coordinates perpendicular to the trajectory (respectively at 
initial and final point), $k$ and $k'$ are initial and final momenta, 
and $\nu$ is a phase (the Maslov index) which, roughly speaking, 
counts points where simple
semiclassical approximation breaks down. Of course, variables
$(\vec{x},\vec{n})$ and  $(\vec{y},\vec{n}')$ are not all independent,
and the sum $(\ref{sumtraj})$ over classical trajectories must take into 
account, for each given starting point and momentum $(\ovx, \ovn)$, all 
classically allowed final points and momenta. 

For 2-dimensional free motion these formulas take especially simple form and
the free Green function (in the units  $m=\dm$ and $\hbar=1$) reads 
(see e.g. \cite{courburneg})
\be
\label{greenf}
G(\ovx,\ovy)=\sum_{tr}
\frac{e^{i k l-i\fpd \nu -i \frac{3\pi}{4}}}{\sqrt{8 \pi k l}},
\end{equation}
where $l$ is the length of the trajectory.

The knowledge of the Green function permits to compute other quantum
quantities. In particular the quantum density of states
\be
\label{defdensite}
d(E)=\sum_{n}\delta(E-E_{n})
\end{equation}
may formally be written by the means of the advanced Green function as
\be
\label{densite}
d(E)=-\frac{1}{\pi}\ \im\int d \ovx \ G(\ovx, \ovx).
\end{equation}
The contribution from very short trajectories  gives the mean level density,
$\bar{d}$, and  the integration of $(\ref{greenf})$ over the whole space
selects periodic orbit
contributions \cite{bb}-\cite{berrytabor} and determines an oscillating
part of level density
\begin{equation}
d(E)=\bar{d}+d^{(osc)}(E).
\end{equation}
For any 2-dimensional billiard  
\begin{equation}
\bar{d}=\frac{\cala}{4\pi}
\end{equation}
where ${\mathcal A}$ is the area of the billiard. The explicit form of 
$d^{(osc)}(E)$ depends of the system considered 
(see e.g. \cite{bb}-\cite{berrytabor}).

Diffractive systems discussed in the paper are characterized by  
the existence of singularities which make the classical motion undetermined. 
Each time a
classical trajectory hits a singularity there is no unique way to continue it.
Quantum mechanics smoothes out these singularities and associates with each
(not too strong) singularity a diffraction coefficient,
$\cald(\vec{n}, \vec{n}')$, which determines a 
scattering  amplitude on this singularity from the initial direction $\vec{n}$ to
the final direction $\vec{n}'$. In the presence of a
singularity the Green function in the whole space has two contributions
\be
G(\ovx, \ovx')=G_0(\ovx, \ovx')+G_d(\ovx, \ovx')
\end{equation}
where $G_0(\ovx, \ovx')$ is the free Green function $(\ref{greenf})$ and 
$G_d(\ovx, \ovx')$ is the contribution of trajectories that hit  
the singularity
\be
\label{gund}
G_d(\ovx, \ovx')=\sum_{\vec{n},\vec{n}'} 
G_{0}(\ovx, (\ovx_{0}, \ovn))\cald(\ovn, \ovn') G_{0}((\ovx_{0}, \ovn'), \ovx')
\end{equation}
where $\cald(\ovn, \ovn')$ defined by Eq.~$(\ref{gund})$ is called 
the diffraction coefficient. Here the quantity $G_{0}(\ovx, (\ovx_{0}, \ovn))$ 
is a contribution to the Green function
from a classical trajectory starting at point $\ovx$ and ending at the singularity
$\ovx_{0}$ with momentum in the direction $\ovn$; $G_{0}((\ovx_{0}, \ovn'), \ovx')$
is  a contribution to the Green function from a classical trajectory starting
at point $\ovx_{0}$ with momentum in the direction $\ovn'$ and ending at point
$\ovx'$. 

For diffractive systems the density of states $(\ref{densite})$ can be written 
as  the sum of three terms \cite{pavloff}, \cite{keller}, \cite{rosenquist}
\be
\label{densitetrois}
d(E)=\bar{d}+d_{p.o.}(E)+d_{d.o.}(E),
\end{equation}
where $\bar{d}$ is the mean level density, $d_{p.o.}$ is the contribution of
periodic orbits without singularity, and the third term, $d_{d.o.}(E)$, is a
contribution from all classical trajectories starting and ending at the
singularity, called diffractive orbits 
\be
d_{d.o.}(E)=\sum_{m=1}^{\infty}\frac{1}{\pi m}\frac{\partial}{\partial
  E}\im\sum_{\vec{n}_{i}, \vec{n}_{j}'}G(\vec{n}_{1}, \vec{n}'_{1})
\cald(\vec{n}'_{1}, \vec{n}_{2})\ldots G(\vec{n}_{m}, \vec{n}'_{m})
\cald(\vec{n}'_{m}, \vec{n}_{1}),
\label{densitedo}
\end{equation}
where $G(\vec{n}, \vec{n}')$ is now the contribution to the Green function from
a classical trajectory starting at the singular point with initial momentum
in direction $\vec{n}$ and ending at it with final momentum in direction
$\vec{n}'$.

For elastic scattering the diffraction coefficient cannot be arbitrary but
has to obey the optical theorem which is the manifestation of the quantum
mechanical unitarity. In the most general form the optical theorem is 
\cite{landau}
\be
\label{unitarite}
\cald(\ovn, \ovn')-\bar{\cald}(\ovn', \ovn)=-\frac{i}{4\pi}
\int\cald(\ovn, \ovn'')\bar{\cald}(\ovn', \ovn'')\ do''
\end{equation}
where $do''$ is the angle giving the direction of $ \ovn''$. 

In particular, a constant diffraction coefficient should have the form
\be
\cald=\frac{\lambda}{1+ \frac{i}{4}\lambda},
\label{dcon}
\end{equation}
with real $\lambda$.

\section{Periodic boundary conditions}\label{section2}

Let us consider a rectangular billiard with sides $a$ and $b$ and with a point-like
scatter at a point $(x_0, y_0)$ inside the rectangle. We always assume that
the ratio $a^2/b^2$ is a `good' (diophantine) irrational number. This
generic requirement is necessary to avoid `accidental' degeneracies of
periodic orbit lengths and under this condition it is possible to prove
\cite{marklof} that the two-point correlation function of such a rectangular
billiard in the ``free'' case (that is without scatter) agrees with 
the Poisson statistics. In this section we
impose periodic boundary conditions.

Fig. \ref{unfoldedperiodic} shows a classical periodic orbit in the rectangle. 
Since
opposite sides of the rectangle are identified, one can unfold a classical
trajectory to a straight line on the plane tiled with rectangles; the images
of the point-like scatter have coordinates $\{x_0+M a, y_0 + N b\}$,
$M,N\in \zm$. 
\begin{figure}[ht]
\begin{center}
\epsfig{file=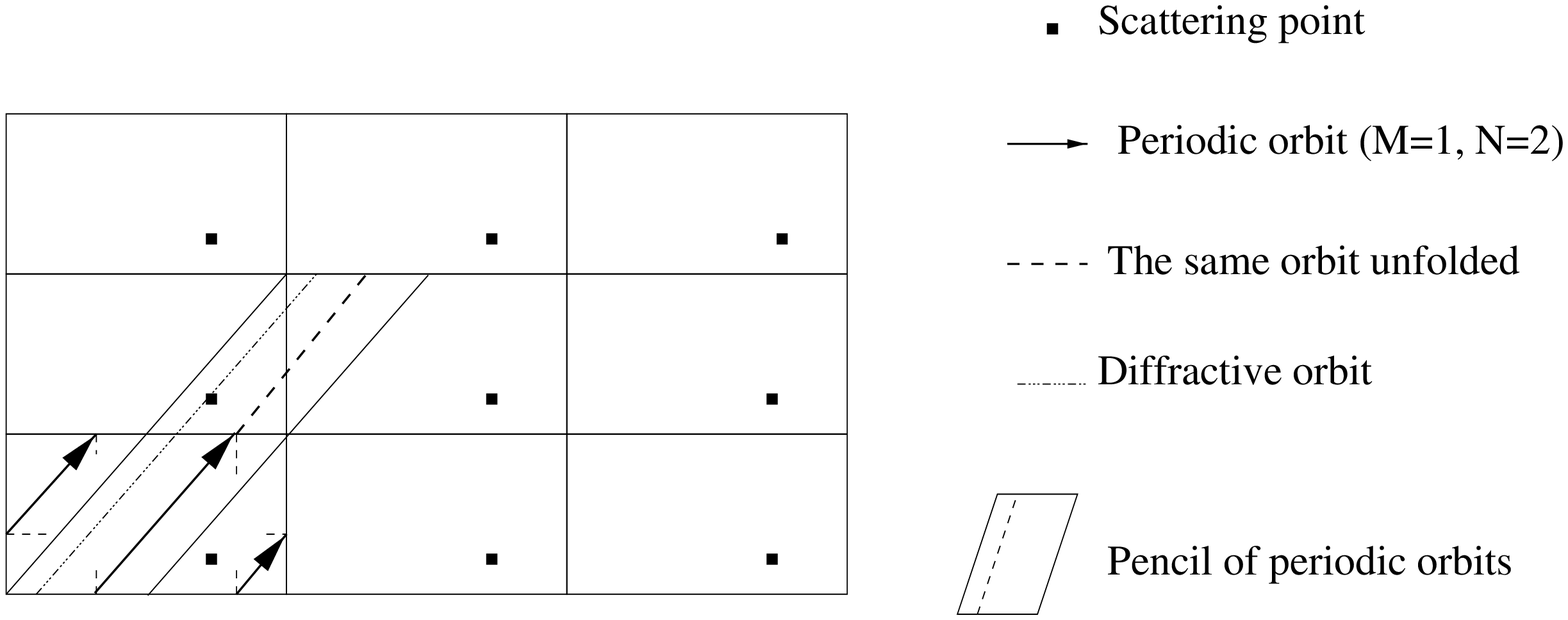,width=13cm}
\end{center}
\caption{An unfolded trajectory in the rectangle}
\label{unfoldedperiodic}
\end{figure}

The unperturbed energy levels of a rectangular billiard with periodic boundary
conditions are
\be
E=(\frac{2\pi }{a}n)^2+(\frac{2\pi }{b}m)^2
\end{equation}
for all integers $m,n=0,\pm 1,\pm 2,\ldots$.

When $mn\neq 0$ energy levels have multiplicity  4. To remove this
trivial degeneracy  it is convenient to consider non-degenerate states only.
Asymptotically it gives the factor  $1/4$ in the periodic orbit density of
states and to be consistent with Ref.~\cite{bogomolny2} and Equations 
$(\ref{densite})$ and $(\ref{gund})$ it is necessary to
multiply the diffraction coefficient by $g=4$ (see below). 

Therefore in our case the  smooth part will be
\begin{equation}
\bar{d}=\frac{{\cala}}{16\pi},
\end{equation}
where ${\mathcal A}=a b$ is the area of the rectangle. The contribution of the
periodic orbits to the density of states is given by \cite{berrytabor}
\begin{equation}
\label{dpo}  
d_{p.o.}(E)=\sum_{p^{+}}
\frac{\cala_p}{2\pi}\frac{1}{\sqrt{8\pi k l_{p}}} 
e^{i k l_{p}-i\frac{\pi}{2}\nu_{p}-i\frac{\pi}{4}} + \cc
\end{equation}
where
\be
\label{lep}
l_p=\sqrt{(M a)^{2}+(N b)^{2}}
\end{equation}
is the length of a periodic orbit. In all integrable
billiards periodic orbits are not isolated but belong to families which
cover an area $\cala_{p}$ (see Fig. \ref{unfoldedperiodic}).
For the rectangular billiard with periodic boundary conditions the area
covered by any family of periodic orbits is $\cala_p={\mathcal A}$.
Since $d_{p.o.}(E)$ corresponds to the non-degenerate density of states, the
summation in $(\ref{dpo})$ is performed over all periodic orbits of
length $l_p$ with $M, N \geq 0$ (which corresponds to divide by 4 the
density $(\ref{densite})$); this is expressed by the index $p^{+}$.

\subsection{Diffractive density of states}

The third term in the density $(\ref{densitetrois})$ is the ``density of
diffractive orbits'' $(\ref{densitedo})$, that is the contribution 
of classical trajectories starting and ending at the
singularity. The diffraction coefficient for our model is a constant
given by $(\ref{dcon})$. The Green functions in
$(\ref{densitedo})$ can be expressed as a sum over all diffractive orbits,
that is over all vectors 
\be
\vec{l}_d= (Ma, Nb)
\label{vecd}
\end{equation}
linking two images of the scatter in the plane tiled with rectangles:
\be
\label{greenperiodic}
G=\sum_{\ovn, \ovn'}
G(\vec{n},\vec{n}')=\sum_{\vec{l}_d}\frac{e^{i k l_{d}-i\fpd \nu_{d} -i \frac{3
      \pi}{4}}}{\sqrt{8 \pi k l_{d}}},
\end{equation}
where $l_d$ is given by expression identical to $(\ref{lep})$
\be
l_d=\sqrt{(M a)^{2}+(N b)^{2}}.
\end{equation}
Note that when $MN\neq 0$ there are $g=4$ diffractive orbits with exactly
the same lengths corresponding to $\pm M$, $\pm N$, As the Green functions
enter in the diffractive trace formula (\ref{densitedo}) always multiplied
by the diffraction coefficient it is convenient to restrict the summation in
(\ref{greenperiodic}) to $M,N\geq 0$ and to multiply the diffraction
coefficient by $g=4$.

Each term in  $(\ref{densitedo})$ involving $m$ Green functions gives the
following  contribution to the density of states
\begin{eqnarray}  
d_{d.o.}^{(m)}(E)&=&\frac{(g\cald)^{m}}{4 i m \pi k }\frac{\partial}{\partial k}
\left(\sum_{\ovn,\ovn'}G(\ovn,\ovn')\right)^{m}+\cc\nonumber\\
&=&\frac{(g\cald)^{m}}{4 i m \pi k }\frac{\partial}{\partial k}
\sum_{\vec{l}_{1},\ldots ,\vec{l}_{m}}
\frac{e^{i k l_1-3 i\pi/4}}{\sqrt{8\pi k l_1}} \ldots 
  \frac{e^{i k l_{m}-3 i\pi/4}}{\sqrt{8\pi k l_{m}}}+\cc
\label{ddodk}  
\end{eqnarray}
$\vec{l}_1,\ldots, \vec{l}_m$ are $m$ diffractive orbits vectors and $l_j$ are their
lengths. The Maslov index in the  periodic case is equal to zero.

The method which permits to treat such sums has been proposed in
\cite{bogomolny2}. Its main content is the existence in the sums over orbits
$\vec{l}_j$ of saddle-point manifolds which correspond to vectors $\vec{l}_j$
almost parallel to a given vector $\vec{L}$ (see Fig.~\ref{saddlepoints}).
\begin{figure}[ht]
\begin{center}
\epsfig{file=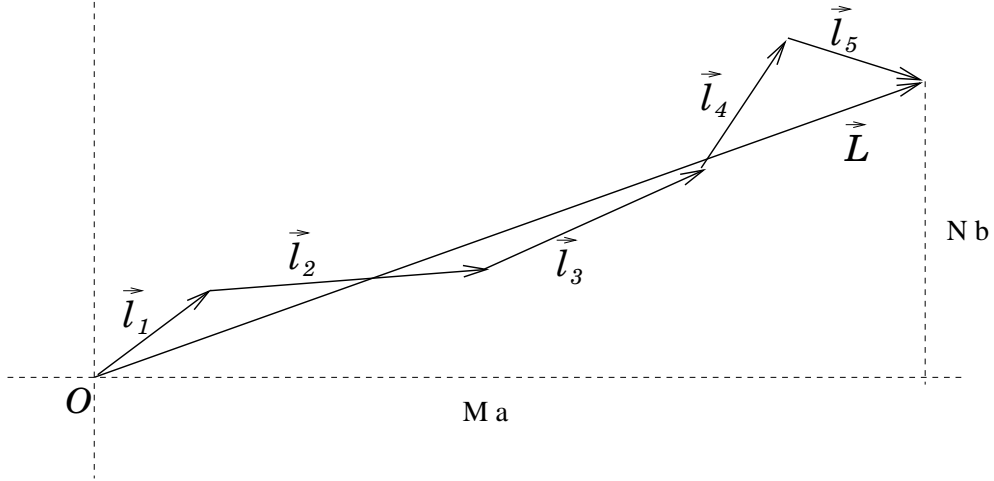,width=13cm}
\end{center}
\caption{A saddle-point contribution to $G^{m}(\vl)$.}
\label{saddlepoints}
\end{figure}

When the $m^{\mbox{th}}$ power of the Green function, $G^m$, is considered 
we can distinguish 
different saddle-point manifolds \cite{bogomolny2}. The
first corresponds to the case when all $m$ vectors
$\vec{l}_1,\ldots,\vec{l}_m$ are close to a unique 
fixed vector $\vec{L}$. The others
are given by contributions where the $\vec{l}_j$, $1\leq j\leq m$, are gathered
into $p$ groups $(2\leq p\leq m)$ with the following properties:

(i) within each group all vectors are almost parallel to a vector
$\vec{L}_i$

(ii) the $p$ vectors $\vec{L}_i$ are quite far from each other (and in
particular their lengths are different). 

Let us first order vectors $\vl_1,\ldots,\vl_p$ by a certain manner, e.g.
according to their lengths 
\be
L_1<L_2<\ldots<L_p.
\end{equation}
Because it is assumed that all $L_j$ are different it is always possible.

Then  following  \cite{bogomolny2} one can write the formal expression 
\be
G^m=\sum_{p=1}^m \sum_{\vl_1,\ldots,\vl_p} 
G_p^m (\vec{L}_1,\ldots , \vec{L}_p),
\end{equation}
where $G$ is given by $(\ref{greenperiodic})$ and
\begin{eqnarray}
\label{gmp}
&&G^{m}_{p}(\vl_1,\ldots ,\vl_p)=\\
&&\sum_{\genfrac{}{}{0pt}{}{m_1+\ldots +m_p=m}{m_i\geq 1}}N(m_1, ..., m_p)
\sum_{\vec{l}_{1},\ldots ,\vec{l}_{m}}
\frac{e^{i k l_1-3 i\pi/4}}{\sqrt{8\pi k l_1}} \ldots 
  \frac{e^{i k l_{m}-3 i\pi/4}}{\sqrt{8\pi k l_{m}}}\nonumber\\
&&\times\delta\left(\vec{l}_{1}+\ldots+\vec{l}_{m_1}-\vl_1\right)\ldots 
\delta\left(\vec{l}_{m_1+\ldots +m_{p-1}+1}+\ldots+\vec{l}_{m}-\vl_p\right)
\nonumber
\end{eqnarray}
for any partition $m_1+\ldots +m_p=m$ of $m$ into a sum of $p$ integers 
It is assumed that vectors $\vec{l}_j$ all have positive
components. $N(m_1, ..., m_p)$ is the number of possible permutations of
the $l_j$ and will be discussed later.

The meaning of this representation is the following (for more details see
\cite{bogomolny2}). The amplitudes corresponding to vectors almost parallel
(i.e. belonging to the same group) should be summed coherently but the ones 
with different vectors $\vec{L}_j$ (that is from different groups)
 are non-coherent. 
Therefore after a smoothing over a small energy window only the square
of the amplitudes of non-coherent contributions will survive.

Let us consider at first the case where $m=2$ and $p=1$ (see
Fig.~\ref{sommegreen})
\be
G^2_1(\vl)=\sum_{\vec{l}_{1},\vec{l}_{2}}
\frac{e^{i k l_1-3 i\pi/4}}{\sqrt{8\pi k l_1}}  
  \frac{e^{i k l_{2}-3 i\pi/4}}{\sqrt{8\pi k l_{2}}}
    \delta (\vec{l}_{1}+\vec{l}_{2}-\vl)
\end{equation}    
(in this case the symmetry factor $N(2)$ is equal to 1).
\begin{figure}[ht]
\begin{center}
\epsfig{file=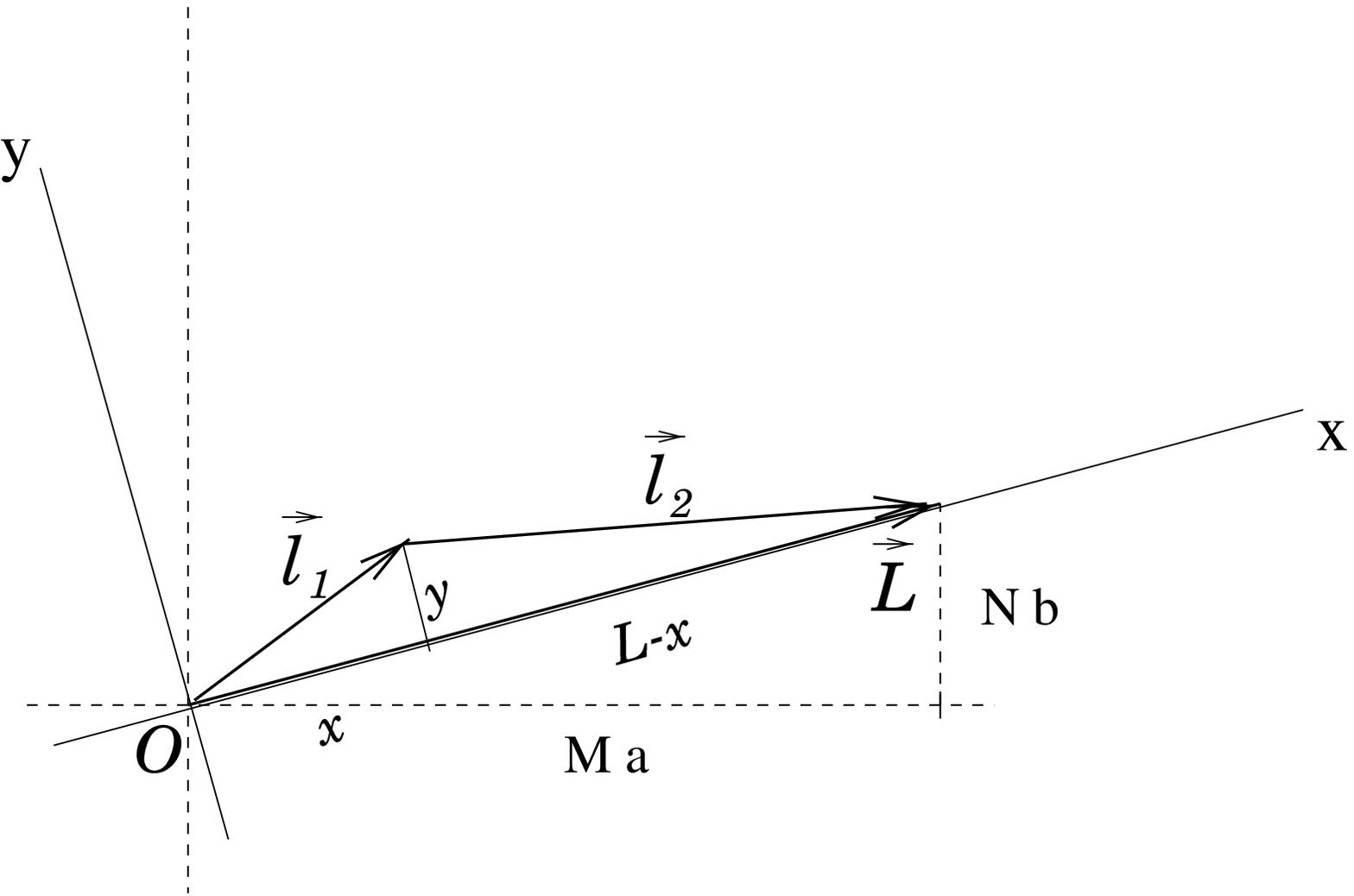,width=10cm}
\end{center}
\caption{Contributions to the product of two Green functions}
\label{sommegreen}
\end{figure}
We have to sum products of 2
Green functions over all vectors  $\vec{l}_1=(m_1a, n_1b)$ and
$\vec{l}_2=(m_2a, n_2b)$ verifying the condition 
\be
\vec{l}_1+\vec{l}_2=\vl.
\end{equation}
It means that we sum over all positive integers $m_1, n_1, m_2,n_2$ such
that this condition is verified. Changing the sum into an integral, we have
to integrate over $m_i$ and $n_i$ ($i=1,2$).
In the coordinates $(x,y)$ where $\vl=(0,L)$ (see Fig. \ref{sommegreen})
we set $\vec{l}_1=(x,y)$, so that
$\vec{l}_2=(L-x,y)$: since $d m_1 d n_1=dxdy/\cala$, we get
\begin{eqnarray}
G^{2}_1(\vl)&=&\int dm_i d n_i \frac{e^{i k l_1-3 i\pi/4}}{\sqrt{8\pi k l_1}}
\frac{e^{i k l_2-3 i\pi/4}}{\sqrt{8\pi k l_2}}
\delta (\vec{l}_{1}+\vec{l}_{2}-\vl)\nonumber\\
&=&\frac{1}{\cala}\int_{0}^{L}dx\int_{-\infty}^{\infty}dy
\frac{e^{i k l_1-3 i\pi/4}}{\sqrt{8\pi k l_1}}
\frac{e^{i k l_2-3 i\pi/4}}{\sqrt{8\pi k l_2}}
\end{eqnarray}
(the condition $0\leq x\leq L$ is necessary to have $\vec{l}_1$ and
$\vec{l}_2$ in the upper right quadrant). In the semi-classical limit where
$k\to \infty$ we use
\be
l_1+l_2\simeq L+\frac{L}{2x(L-x)}y^2
\end{equation}
and after performing the integrals
\be
G^{2}_1(\vl)=\frac{L}{2ik\cala}\frac{e^{i k L-3 i\pi/4}}{\sqrt{8\pi k L}}.
\end{equation}
More generally this method can be iterated to get
\be
\label{gmvl}
G^{m}_1(\vl)=\left(\frac{L}{2ik\cala}\right)^{m-1}\frac{1}{(m-1)!}
\frac{e^{i k L-3 i\pi/4}}{\sqrt{8\pi k L}}.
\end{equation}
The saddle-point manifold in this case is the space of sets of $m$ vectors 
almost parallel to $\vl$ 
such that their sum is equal to $\vl$ (see Fig. \ref{saddlepoints}).
This result can also be obtained almost without calculations by using
Stokes' theorem. 

It may be noticed that the part of the Green function corresponding to the sum 
of contributions $G^{m}_1$,
$\sum_{m=1}^{\infty} (g\cald)^{m-1}G^m_1(\vl)$,
is the perturbation series expansion of the Green function and from
(\ref{gmvl}) one gets
\be
\sum_{m=1}^{\infty} (g\cald)^{m-1}G^m_1(\vl)=\exp (\frac{\rho g\cald}{2 i k}L)G(L)
\label{sum}
\end{equation}
where $\rho=1/\cala$ is the density of scatters and 
\be
G(L)=\frac{e^{i k L-3 i\pi/4}}{\sqrt{8\pi k L}}
\label{GL}
\end{equation}
is the usual contribution to the free Green function of a classical trajectory
of length $L$. The result $(\ref{sum})$ is exactly the exponential attenuation
one expects for the coherent propagation
of a particle in a diffractive medium with coefficient of
diffraction $\cald$. 

The terms $G^m_p(\vl_1,\ldots ,\vl_p)$ can be computed (see \cite{bogomolny2})
by the same me\-thod by considering all possible partitions of the $m$
vectors  into   $p$ groups of almost parallel vectors. But in this case a 
combinatorial factor which counts the exact degeneracies has to be taken 
into account. For a given partition of $m$ into a sum of $p$ integers 
$m_1,\ldots ,m_p$ 
\be
m_1+\ldots +m_p=m
\end{equation}
we obtain a saddle-point manifold in the sum $(\ref{gmp})$ by
choosing $m_1$ vectors parallel to $\vec{L}_1$, $m_2$ vectors parallel to 
$\vec{L}_2$,
and so on. But the numbers $m_i$ do not fix a diffractive trajectory uniquely.
Trajectories built from the same set of primitive diffractive  orbits
connected in a different order have exactly the same lengths and numbers $m_i$.

All such trajectories correspond to one of the permutation of the following
sequence of $p$ symbols
\be
\underbrace{1\ldots 1}_{m_1}\; \underbrace{2\ldots 2}_{m_2}\; 
\ldots \underbrace{p\ldots p}_{m_p}
\label{sequence}
\end{equation}
where it is assumed that in the sequence there exist $m_1$ symbols of
type 1, $m_2$ symbols of type 2, etc, and finally $m_p$ symbols of type $p$.
Each symbol $j$ corresponds to a different vector $\vec{L}_j$ defining
a certain primitive diffractive orbit and a permutation of the sequence 
$(\ref{sequence})$ describes how a
composite diffractive orbit is built. 

For example, at Fig.~\ref{degeneracies} trajectories with $m=4$ and $p=2$
are de-pictured. Fig.~\ref{degeneracies}a corresponds to all partitions with
$m_1=3$ and $m_2=1$ 
\be
1112\;\;1121\;\;1211\;\;2111
\end{equation}
and Fig.~\ref{degeneracies}b represents partitions with $m_1=m_2=2$  
\begin{equation}
1122\;\;1212\;\;1221\;\;2211\;\;2121\;\;2112.
\label{example}
\end{equation}
At these figures trajectories of types 1 and 2 are represented respectively 
by horizontal and vertical lines.
\begin{figure}[ht]
\begin{center}
\epsfig{file=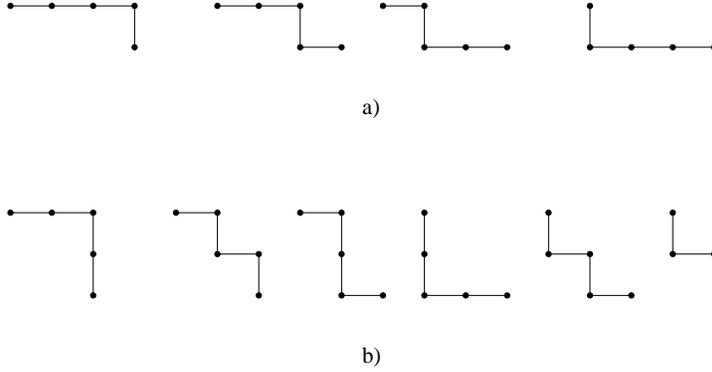,width=10cm}
\end{center}
\caption{Trajectories with $m=4$ and $p=2$ with exactly the same length.
a) $m_1=3, m_2=1$; b) $m_1=2, m_2=2$.  }
\label{degeneracies}
\end{figure}
Since the diffraction coefficient is a constant the  corresponding 
saddle-point manifolds (consisting of
trajectories close to these exactly degenerate ones) all give the same contribution
to the density $d_{d.o.}$ ; therefore 
the contribution from one saddle point has to be multiplied by their number
which is the number of partitions of $m$ into $p$ groups with $m_i$ elements
in the $i^{th}$ group  
\be
\label{facteurcombi}
N(m_1,\ldots,m_p)=\frac{(m_1+\ldots+m_p)!}{m_1 ! \ldots m_p !}.
\end{equation}
These exact multiplicities of diffraction orbits clearly demonstrate the
difference between scattering on regular and random arrays. For a random
distribution of scattering points all trajectories as at
Fig.~\ref{degeneracies} will have different lengths but, because in our model
all points of diffraction are copies of the same scattering point, these 
trajectories have equal lengths. Finally
\begin{equation}
\label{gmvlp}
G^{m}_p(\vl_1,\ldots ,\vl_p)= 
\sum_{\genfrac{}{}{0pt}{}{m_1+\ldots +m_p=m}{m_i\geq 1}}
\frac{m!}{m_1 ! \ldots m_p !}
\prod_{i=1}^{p}\left(\frac{L_i}{2ik\cala}\right)^{m_i-1}
\frac{G(L_i)}{(m_i-1)!},
\end{equation}
where $G(L)$ is defined in (\ref{GL}).

According to Equation (\ref{ddodk}), the density of diffractive orbits is
the derivative of the powers of the Green function. To get the leading term
in the semiclassical approximation it is necessary to
differentiate only the exponent and the diffractive density of states takes
the form
\be
d_{d.o.}(E)=\sum_{p=1}^{\infty}
\sum_{L_1<\ldots<L_p}d_{d.o.}^{(p)}(L_1,\ldots,L_p)G(L_1)\ldots G(L_p)+c.c.,
\end{equation}
where $\vl_1,\ldots ,\vl_p$ are summed over the upper-right quadrant
and
\begin{eqnarray}
&&d_{d.o.}^{(p)}(L_1,\ldots,L_p)=\frac{L_1+\ldots+L_p}{4\pi k} 
\sum_{m=p}^{\infty}\frac{1}{m}
\sum_{\genfrac{}{}{0pt}{}{m_1+\ldots +m_p=m}{m_i\geq 1}}
R(m_1,\ldots,m_p)\nonumber\\
&&\times (\frac{L_1}{2ik\cala})^{m_1-1}\frac{1}{(m_1-1)!}\ldots
(\frac{L_p}{2ik\cala})^{m_p-1}\frac{1}{(m_p-1)!},
\label{ddomp}
\end{eqnarray}
where 
\be
\label{rmp}
R(m_1,\ldots,m_p)=(g\cald)^{m_1+\cdots+m_p}
\frac{(m_1+\ldots+m_p)!}{m_1!\ldots m_p!};
\end{equation}
$d_{d.o.}^{(p)}(L_1,\ldots,L_p)$ is the contribution of all diffractive 
trajectories (like those at
Fig.~\ref{degeneracies}) built from $m_1$ vectors
$\vl_1$, $m_2$ vectors $\vl_2$, etc.

The term with $p=1$ is
\be
d_{d.o.}^{(1)}(L)=\frac{i\cala}{2\pi}\sum_{m=1}^{\infty}
\frac{1}{m!}(\frac{g\cald L}{2ik\cala})^m=\frac{i\cala}{2\pi}
(\exp (\frac{g\cald}{2ik\cala}L)-1).
\end{equation}
Since the contribution of periodic orbits
(\ref{dpo}) can be rewritten
\be
\sum_{L}\frac{i\cala}{2\pi}G(L),
\end{equation}
one gets, adding all contributions,
\be
d^{(osc)}(E)=\sum_{p=1}^{\infty}
\sum_{L_1<\ldots<L_p}d_{p}(L_1,\ldots,L_p)G(L_1)\ldots G(L_p)+c.c.,
\label{dmain}
\end{equation}
where
\be
d_1(L)=\frac{i\cala}{2\pi}\exp (\frac{g\cald}{2ik\cala}L),
\label{d1}
\end{equation}
and $d_{p}(L_1,\ldots,L_p)=d_{d.o.}^{(p)}(L_1,\ldots,L_p)$ for $p\geq 2$.

\subsection{Two-point correlation form factor}

The  2-point correlation function is related with the level density by the usual
formula
\be
\label{corr}
R_{2}(\eps)=\left< d(E+\frac{\eps}{2})\ d(E-\frac{\eps}{2}) \right>,
\end{equation}
where the brackets denote an energy averaging around $E$ over an interval of 
energy much larger than the mean level spacing $1/\bar{d}$  and much smaller 
than energy $E$.

The two-point correlation form factor is defined as the Fourier transform of 
the connected part of $R_{2}(\epsilon)$ :
\be
\label{defk}
K(\tau)=\int_{-\infty}^{\infty}\frac{d\eps}{\bar{d}}
\left<d^{(osc)}(E+\frac{\eps}{2})\ d^{(osc)}(E-\frac{\eps}{2}) \right> 
e^{2 i \pi \bar{d}\eps\tau},
\end{equation}
(the factors are chosen so that $\tau$ and $K(\tau)$ are dimensionless).

The advantage of the representation (\ref{dmain}) is that it is a sum
over saddle-points (each set of vectors 
$(\vec{L}_1,...,\vec{L}_p)$ for $p\geq 1$ is a saddle-point)
which are uncorrelated, so we can
apply the usual diagonal approximation to this sum: in the cross product
(\ref{defk}) only squares of terms corresponding to a given saddle-point 
will give a contribution. 

Using the expansion of momentum $\sqrt{E+\eps}\simeq\sqrt{E}+\eps/(2\sqrt{E})$ we 
get
\be
R_2(\eps)=\sum_{p=1}^{\infty}\frac{1}{(8\pi k)^{p}}
\sum_{L_1<\ldots<L_p}|d_p(L_1,\ldots, L_p)|^2
\frac{e^{i\eps (L_1+\ldots+L_p)/(2k)}}{L_1\cdots L_p}+\cc,
\label{r2final}
\end{equation}
where $d_{p}(L_1,\ldots,L_p)$ are defined in (\ref{dmain}).

As all quantities depend only of the lengths of vectors $\vec{L}_i$ and
$L_i \rightarrow \infty$ one can substitute the summation over $\vec{L}_i$
by the integration with the density 
\be
\label{rhol}
\rho(l)=\int_{0}^{\infty}dMdN\delta(l-\sqrt{(Ma)^2+(Nb)^2})=
\frac{\pi l}{2 \cala}.
\end{equation}
(Remind that we consider vectors in the upper-right quadrant with positive 
 $M,N$.)

Because
\be
\sum_{L_1<\ldots <L_p}=\frac{1}{p!}\sum_{L_1,\ldots ,L_p}
\end{equation}
for any symmetric summand one obtains 
\begin{eqnarray}
R_2(\eps)&=&\sum_{p=1}^{\infty}\frac{1}{p!}
\int_0^{\infty}\frac{dL_1}{16\cala k} \ldots
\int_0^{\infty}\frac{dL_p}{16\cala k}  |d_p(L_1,\ldots, L_p)|^2
e^{i\eps (L_1+\ldots+L_p)/(2k)}\nonumber\\
&+&\cc.
\label{rp}
\end{eqnarray}
The two-point correlation form factor has the similar form
\be
K(\tau)=\sum_{p=1}^{\infty}K_p(\tau)
\end{equation}
where
\begin{eqnarray}
K_p(\tau)&=&\frac{4\pi k}{\bar{d} p!}
\int_0^{\infty}\frac{dL_1}{16\cala k} \ldots
\int_0^{\infty}\frac{dL_p}{16\cala k}  |d_p(L_1,\ldots, L_p)|^2
\nonumber\\
&\times& \delta (L_1+\ldots+L_p-4\pi \bar{d}k\tau),
\label{kp}
\end{eqnarray}
and $\bar{d}=\cala/16\pi$.

The term with $p=1$ is especially simple (see (\ref{d1}))
\be
\label{kpp1}
K_1(\tau)=|\exp (-i\frac{\cald g}{8}\tau)|^2.
\end{equation}
Using the optical theorem (\ref{unitarite}) this expression can be rewritten in
the following form
\be
\label{kpp}
K_1(\tau)=\exp (-\frac{|\cald|^2 g}{16}\tau).
\end{equation}
Substituting (\ref{ddomp}) into (\ref{kp})  one obtains the
contributions from terms with $p\geq 2$ 
(we define $\tau_i$ by $L_i=4\pi k\bar{d}\tau_i$)
\begin{eqnarray}
&&K_p(\tau)=\frac{\tau^2}{p!}
\int_0^{\infty}d\tau_1\ldots\int_0^{\infty}d\tau_p 
\sum_{m,n\geq p}
\sum_{\genfrac{}{}{0pt}{}{m_1+\ldots +m_p=m}{n_1+\ldots +n_p=n}}
\frac{(8i)^{-m}(-8i)^{-n}}{mn }
\nonumber\\
&&\times R(m_1,\ldots,m_p) \bar{R}(n_1,\ldots,n_p)\nonumber\\
&&\times 
\left [\prod_{i=1}^p
  \frac{\tau_i^{m_i+n_i-2}}{(m_i-1)!(n_i-1)!}\right ] 
  \delta (\tau_1+\ldots \tau_p-\tau)
\label{kpmain}  
\end{eqnarray}
where all $m_i, n_i\geq 1$.
The remaining integral has the form
\be
J(\tau)=\int_{0}^{\infty}d \tau_1\ldots d \tau_p\ 
\tau_1^{m_1+n_1-2}\ldots \tau_p^{m_p+n_p-2}
\delta(\tau_1+\ldots +\tau_p -\tau)
\end{equation}
and can easily be computed by the Laplace transform
\be
\label{tfj}  
J(\tau)=
\frac{\tau^{m+n-p-1}}{(m+n-p-1)!}\prod_{r=1}^{p}(m_r+n_r-2)!.
\end{equation}
With Eqs. $(\ref{kpmain})-(\ref{tfj})$ one obtains
the contribution to  the form factor
\begin{eqnarray}
K_p(\tau)&&=\frac{1}{p!}\sum_{m,n\geq p}
\sum_{\genfrac{}{}{0pt}{}{m_1+\ldots +m_p=m}{m_i\geq 1}}
\sum_{\genfrac{}{}{0pt}{}{n_1+\ldots +n_p=n}{n_i\geq 1}}
\frac{1}{mn} R(m_1,\ldots,m_p)R(n_1,\ldots,n_p)\nonumber \\
&&\times \prod_{r=1}^{p}\left[\frac{(m_r+n_r-2)!}{(m_r-1)!(n_r-1)!}\right]
\frac{(8i)^{-m}(-8i)^{-n}}{(m+n-p-1)!}\tau^{m+n-p+1}.
\label{kpourp}  
\end{eqnarray}
Adding all factors together we finally get the complete
perturbation series expansion of the two-point correlation form factor for
rectangular billiard with a delta-function impurity and periodic boundary
conditions
\begin{eqnarray}
\label{kmainper}
K(\tau)&=&e^{-|\cald|^2 g\tau/16}\\
&+&\sum_{p=2}^{\infty}\frac{1}{p!}\sum_{m,n\geq 0}A_{mnp}  
\left(\frac{-i\cald g}{8}\right)^{m+p}
\left(\frac{i\bar{\cald} g}{8}\right)^{n+p}\tau^{m+n+p+1}\nonumber
\end{eqnarray}
where we have defined the rational numbers
\be
\label{amnper}
A_{mnp}= \frac{(m+p-1)!(n+p-1)!}{(m+n+p-1)!}
\sum_{m_i, n_j\geq 0}
\prod_{i=1}^{p}\left[\frac{C_{m_i+n_i}^{m_i}}{(m_i+1)!(n_i+1)!}\right]
\end{equation}
(the sum is taken over all non-negative integers $m_i$ and $n_j$
verifying $m_1+\cdots+m_p=m$ and $n_1+\cdots+n_p=n$). Remind that in our
case $g=4$. This result coincides
with the expansion of the exact formula derived in \cite{gerland}
(see Appendix).

When $\tau\rightarrow 0$, $K(\tau)\rightarrow 1$ as  expected for generic 
integrable billiards \cite{berry}.

The special case $\cald=-4i$ obtained for $\lambda\to\infty$ in $(\ref{dcon})$
corresponds to spectral statistics of the star
graphs  \cite{Berko} where Eq.~(\ref{kmainper}) for this value of the
diffraction coefficient yields
\be
K(\tau)=e^{-4\tau}+\sum_{p=2}^{\infty}\frac{(-2)^{p-1}}{p!}
\sum_{m,n\geq 0}A_{mnp}  (-2\tau)^{m+n+p+1}
\end{equation}
and can be derived by another method \cite{berkolaiko}.

\section{Dirichlet boundary conditions}\label{section3}

Let us  consider now the case where we impose  Dirichlet
conditions on boundary of a rectangular billiard  which means that the 
eigenfunctions 
of the quantum problem vanish  on the boundary. It is well known that
for this boundary conditions the classical motion corresponds to the specular
reflection on the boundary.
For example, Fig.~\ref{unfoldeddirichlet} shows a periodic
orbit in the rectangle.
\begin{figure}[ht]
\begin{center}
\epsfig{file=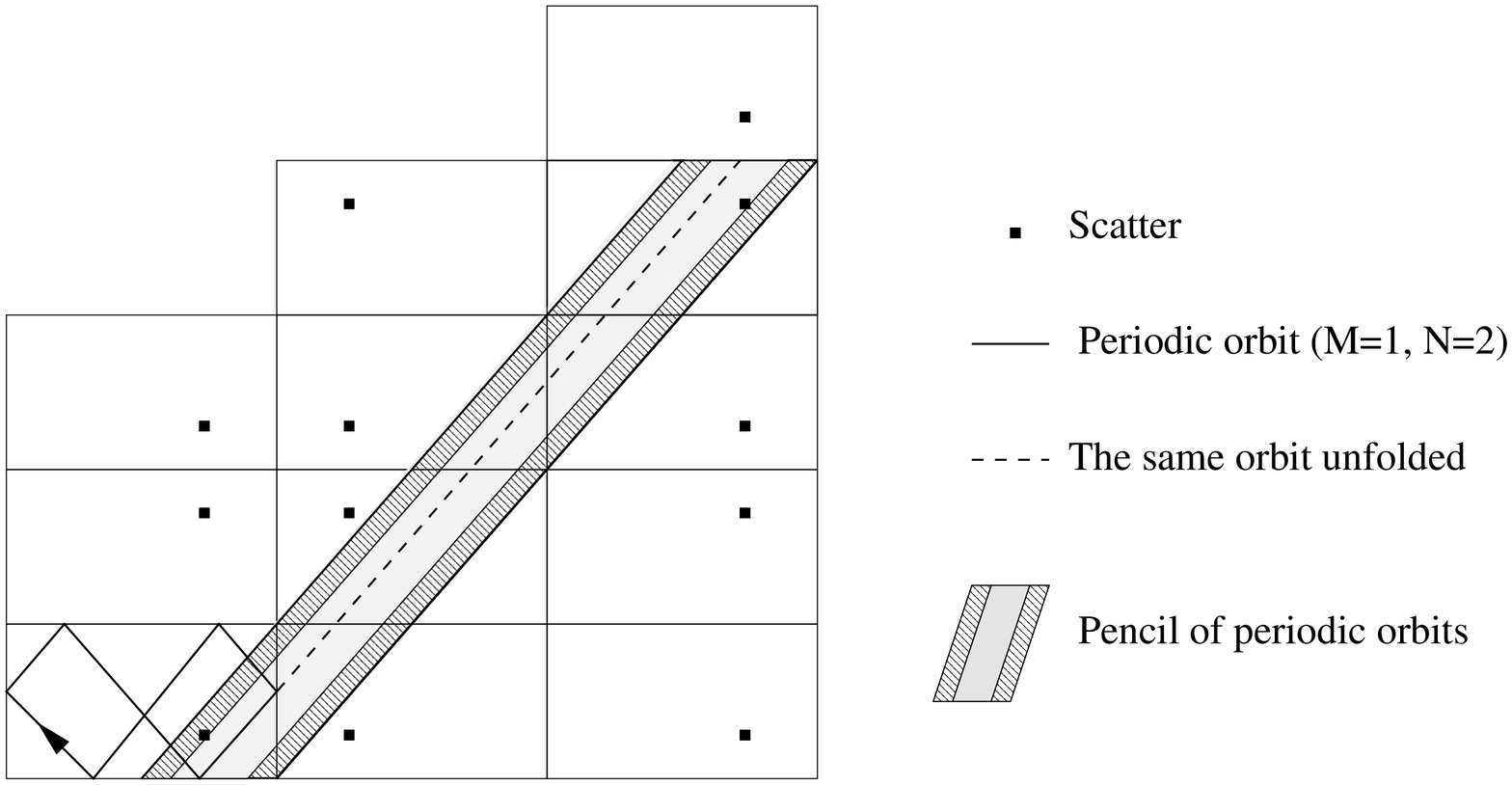,width=13cm}
\end{center}
\caption{An unfolded trajectory in the rectangle}
\label{unfoldeddirichlet}
\end{figure}
One can unfold the trajectory to a straight line on
the plane tiled with rectangles; the images of the point-like scatter have
coordinates $\{(\eps_1 x_0+2 M) a, (\eps_2 y_0 + 2 N) b\}$, where as before
$(x_0,y_0)$ are coordinates of the scatter, $M,N$ are in $\zm$ and $\eps_{1}$
and $\eps_2$  take the values $\pm 1$.

The density of states still contains three terms
\begin{equation}
d(E)=\bar{d}+d_{p.o.}(E)+d_{d.o.}(E)
\end{equation}
but in this case the sum $(\ref{defdensite})$ over the eigenvalues is not
degenerate any more. Therefore the smooth part is 
\begin{equation}
\bar{d}=\frac{{\mathcal A}}{4\pi},
\end{equation}
where ${\mathcal A}=a b$ is the area of the rectangle. The density of
periodic orbits is given by an expression similar to $(\ref{dpo})$ : 
\begin{equation}
\label{dpod}  
d_{p.o.}(E)=
g\sum_{p^{+}}\frac{\cala}{2\pi}\frac{1}{\sqrt{8\pi k l_{p}}} 
e^{i k l_{p}-i\frac{\pi}{2}\nu_{p}-i\frac{\pi}{4}} + \cc,
\end{equation}
where the summation is still performed over all periodic orbits of
length $l_p$ with $M, N \geq 0$. But here
\be
\label{lepdir}
l_p=\sqrt{(2 M a)^{2}+(2 N b)^{2}}
\end{equation}
and for the rectangular billiard with Dirichlet boundary conditions
the multiplicity of periodic orbit lengths is $g=4$ for $MN\neq 0$.

The density of diffractive orbits is given by the usual formula $(\ref{densitedo})$ 
with diffraction coefficients $(\ref{dcon})$. The
Green function in $(\ref{densitedo})$  can be expressed as a sum over all 
diffractive orbits,
that is all vectors linking two images of the scatter in the plane tiled
with rectangles:
\be
\label{greendirichlet}
G=\sum_{\ovn, \ovn'}
G(\vec{n},\vec{n}')=\sum_{\vec{l}_{d}}\frac{e^{i k l_{d}-i\fpd \nu_{d} -i \frac{3
      \pi}{4}}}{\sqrt{8 \pi k l_{d}}},
\end{equation}
where the sum is taken over $M, N \in\zm$ since there is no degeneracy 
of the energy levels any
more; $l_d$ is the length of a
trajectory going from $(x_0,y_0)$ to an image of the scatter. 

As in the periodic case we could restrict our investigation to $M,N\geq 0$
but now we have to consider the 16 vectors
whose lengths are almost equal. Those are  vectors
linking $(x_0, y_0)$ to $(\pm 2 M a \pm x_0, \pm 2 N b \pm y_0)$. The
lengths of these orbits are slightly different and can be expanded to the first
order in $1/M, 1/N$. The vector linking $(x_0, y_0)$ to
$(2 M a + \eps_1 x_0, 2 N b +\eps_2 y_0)$ (with $\eps_i=\pm 1$) has a length
\begin{eqnarray}
l_d&=&\sqrt{(2 M a + (\eps_1-1) x_0)^{2}+(2 N b +(\eps_2-1)y_0)^{2}}\nonumber\\
&\simeq&l_p+ (\eps_1-1) \frac{2 M a x_0}{l_p}+ (\eps_2-1)\frac{2 N b y_0}{l_p},
\label{expansionli}
\end{eqnarray}
where $l_p$ is given by $(\ref{lepdir})$. Taking
into account that for this trajectory and the Dirichlet boundary conditions
the phase $\exp (-\dm i\pi \nu_{d})=\eps_1\eps_2$
the terms corresponding to a given $(M,N)$ in the Green function
$(\ref{greendirichlet})$  can be gathered together. There are 16 type of
such orbits corresponding to $\eps_1, \eps_2=\pm 1$ and all possible changes
of signs of $M$ and $N$ (see Fig. \ref{trajdir}). Their sum is
\be
\sum_{i=1}^{16}e^{i k l_{i}-i\fpd \nu_{i}}=
16e^{i k l_{p}} \sin^2 (\phi_1)\sin^2 (\phi_2),
\end{equation}
where
\begin{eqnarray}
\phi_1=2 k\frac{M}{l_p} x_0 a=k x_0 \cos\varphi,
\label{phi1}\\
\phi_2= 2 k\frac{N}{l_p} y_0 b=k y_0 \sin\varphi
\label{phi2}
\end{eqnarray}
and $\varphi$ is the angle between the periodic orbit vector $\vl$ and the
horizontal side of the rectangle. 
For Neumann boundary conditions $\sin $ in the above expression should be
substituted by $\cos$. 

Therefore the difference between billiards with periodic boundary conditions
considered in the previous Section and  the Dirichlet case is the
factor $r(\varphi)$ in $(\ref{greendirichlet})$
\be
\sum_{\ovn, \ovn'}
G(\vec{n},\vec{n}')=g\sum_{p}\frac{e^{i k l_{p} -i \frac{3
      \pi}{4}}}{\sqrt{8 \pi k l_{p}}}r(\varphi),
\end{equation}
where the sum is taken over orbits with $M,N\geq 0$ and $g=4$. Here we
have set
\be
\label{rvarphi}
r(\varphi)=4\sin^{2}\phi_1(\vp)\sin^{2} \phi_2(\vp).
\end{equation}
Compared to the problem with periodic boundary conditions, the Green
function $(\ref{gmp})$ is modified by a product of factors
$r(\varphi_i)$, where
$\varphi_i$ is the angle between the periodic orbit $\vl_i$ and the
horizontal side of the rectangle. One can also leave the Green function
unchanged but substitute the constant diffraction coefficient $\cald$ by an 
effective coefficient, $S(\varphi)$, defined by
\begin{equation}
S(\varphi)=\cald r(\varphi),
\end{equation}
and taken at the saddle-point. The quantities
$G^{m}_p(\vl_1,\ldots ,\vl_p)$ defined by Eq. $(\ref{gmp})$ are unchanged, 
and performing the same steps as
in the previous Section one obtains the formula for the density of states
quite similar to Eqs.~(\ref{ddomp}), (\ref{dmain})
\be
d^{(osc)}(E)=\sum_{p=1}^{\infty}d_p(L_1,\ldots,L_p)G(L_1)\ldots G(L_p)+\cc,
\label{dmaindir}
\end{equation}
where
\be
d_1(L)=\frac{i\cala}{2\pi}\exp\left(g\cald r(\varphi) L/(2ik\cala)\right)
\label{d1dir}
\end{equation}
and $d_p(L_1,\ldots,L_p)$ is given by Eq.~$(\ref{ddomp})$ with
\be
\label{Rdircst}
R(m_1,\ldots,m_p)=(g\cald)^{m_1+\cdots+m_p}
\frac{(m_1+\ldots+m_p)!}{m_1!\ldots m_p!}r(\vp_1)^{m_1}...r(\vp_p)^{m_p}
\end{equation}
The next steps of the computation of the two-point correlation form factor
are almost the same as above. The only difference with the periodic case is
the necessity to know the density of periodic orbits, $\rho(l,\varphi)$, 
with fixed length and angle. Like in (\ref{rhol}) one gets
\begin{equation}
\rho(l,\varphi)=\frac{l}{4\cala}.
\end{equation}
In the semi-classical limit $k\to \infty$ under the assumption that the
ratio $(x_0a) /(y_0 b)$ is an irrational number the factors $\phi_1$ and
$\phi_2$ in (\ref{phi1}) and (\ref{phi2}) are equivalent to independent
random variables  uniformly distributed between $0$ and
$\dm\pi$. It means that
\be
<f>=\lim_{k\to \infty}\frac{2}{\pi}\int_0^{\pd }f(\phi_1(\vp),\phi_2(\vp))d\vp=
\frac{4}{\pi^2}\int_0^{\pd }\int_0^{\pd }f(\phi_1,\phi_2)d\phi_1d\phi_2
\end{equation}
for any smooth function $f(\phi_1,\phi_2)$.

The integration over the angle $\varphi$ therefore gives for any
integer $n$,
\begin{equation}
<r(\varphi)^n>=\frac{4}{\pi^2}
\int_{0}^{\pd }\int_{0}^{\pd }d\phi_1d\phi_2
\ r^n(\phi_1,\phi_2),
\label{phimean}
\end{equation}
where we have  set
\be
\label{valeurder}
r(\phi_1,\phi_2)=4\sin^{2}\phi_1\sin^{2}\phi_2.
\end{equation}
Of course, the introduction of $r(\phi_1,\phi_2)$ is just a
convenient way to perform the generalized diagonal approximation where  
one takes into account not only trajectories with exactly the same lengths but
also ones which have equal lengths up to the first order expansion as in
Eq.~(\ref{expansionli}) (see \cite{bogomolny2}). 

The only necessary integral is thus 
\be
<r^{n}>=\frac{(C_{2n}^n)^2}{4^n}=\left(\frac{(2 n)!}{2^n (n!)^{2}}\right)^{2}.
\label{meanvalue}
\end{equation}
The final expression for the 2-point correlation form factor
of the rectangular billiard with Dirichlet boundary conditions is similar to
Eq.~(\ref{kmainper})
\begin{eqnarray}
\label{ffdir}  
K(\tau)&=&\left<e^{-|\cald|^2 g r \tau/16}\right>\\
&+&\sum_{p=2}^{\infty}\frac{1}{p!}\sum_{m,n\geq 0}A_{mnp}  
\left(\frac{-ig\cald }{8}\right)^{m+p}
\left(\frac{i\bar{g\cald} }{8}\right)^{n+p}\tau^{m+n+p+1}\nonumber  
\end{eqnarray}
where 
\be
A_{mnp}= \frac{(m+p-1)!(n+p-1)!}{(m+n+p-1)!}
\sum_{m_i, n_j\geq 0}
\prod_{i=1}^{p}\left[\frac{C_{m_i+n_i}^{m_i}<r^{m_i+n_i+2}>}
{(m_i+1)!(n_i+1)!}\right].
\label{amndir}
\end{equation}
The sum here is taken as before over all non-negative $m_i$ and $n_j$ verifying
$m_1+\cdots+m_p=m$ and $n_1+\cdots+n_p=n$) and the mean value $<\ldots>$ is
evaluated in (\ref{meanvalue}). Note that only the value of $A_{mnp}$ differs
between the periodic case $(\ref{kmainper})$ and Eq.~$(\ref{ffdir})$; and 
if $r$ is set equal to $1$, $(\ref{amndir})$
reduces to $(\ref{amnper})$. In Appendix it is demonstrated that this
expansion agrees with the exact results obtained in Ref.~\cite{gerland}.

\section{Non-constant diffraction coefficient}\label{section4}

In the previous Sections we have assumed that the diffraction coefficient
that appears in the expression of the density of diffractive orbits
$(\ref{densitedo})$ is a constant $(\ref{dcon})$ for a point-like scatter. 
In this Section we consider the general case when this diffraction coefficient
depends on scattering  angles. 

Let a trajectory hit the singularity with momentum in the direction $\ovn$ and 
leave it with momentum in the direction $\ovn'$. The diffraction
coefficient $\cald(\ovn,\ovn')$ is a certain function obeying the optical
theorem (\ref{unitarite}).

\subsection{Periodic boundary conditions}

Trajectories with equal length play a very important role in the computation 
of the form factor. In the rectangular billiard with periodic boundary
conditions the lengths of periodic and diffractive orbits are given by
\be
l=\sqrt{(Ma)^2+(Nb)^2}
\end{equation}
and orbits with $M$, $N$ of different signs are degenerated though
geometrically they are different. When the
diffraction coefficient is a constant all such trajectories give the same
contribution but when it depends on initial and final directions their
contribution will be different.

In the rectangular billiard  orbits that hit or leave the singularity with any
of the 4 angles 
\be
g_\alpha(\vp)= \vp,\;-\vp,\; -\pi+\vp,\;\pi-\vp
\label{angles}
\end{equation}
all have the same lengths (see Fig. \ref{trajper}). 
\begin{figure}[ht]
\begin{center}
\epsfig{file=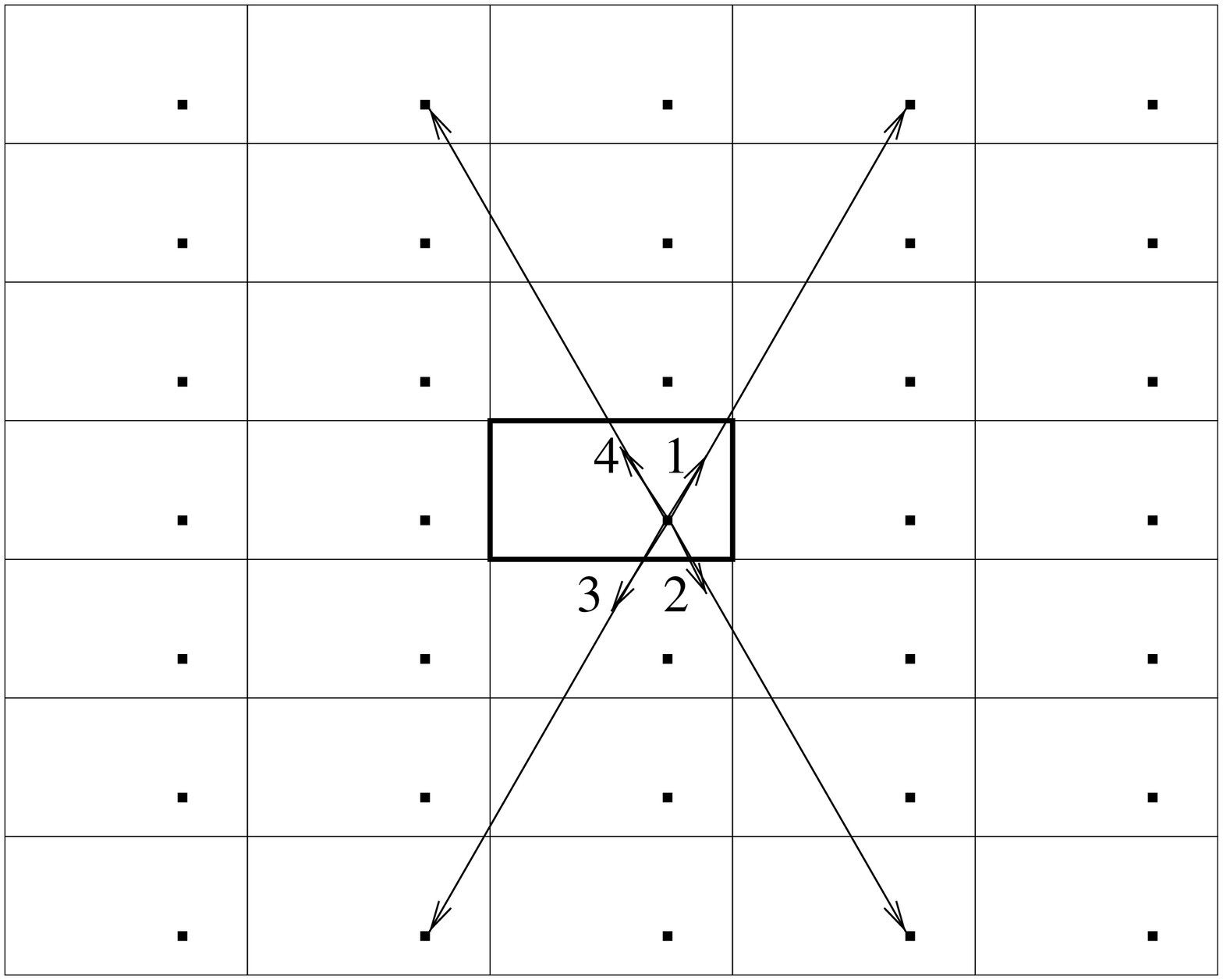,width=10cm}
\end{center}
\caption{Four trajectories of same length in the rectangle with periodic
  boundary
  conditions}
\label{trajper}
\end{figure}
Let us assume that $0\leq \vp<\pd $ and label these angles
respectively by $\alpha=1$, 2, 3 and 4. To take into account
 trajectories with equal length
it is convenient to label them by 2 quantities $(\varphi,\alpha)$
where the angle $\varphi$  is uniquely related with the length of the trajectory 
and belongs to the upper-right quadrant, $0\leq \vp<\pd $, and the index 
$\alpha=1,2,3,4$  describes the form of the trajectory and determines which 
of the 4 
angles (\ref{angles}) is the geometrical  angle between the trajectory and 
the horizontal.

Let us denote by $D_{\alpha \beta}(\vp, \vp')$ the 
diffraction coefficient  $\cald(g_\alpha(\vp), g_\beta(\vp'))$ corresponding
to a scattering process where the trajectory arrives with angle
{\it opposite to} $(\vp,\alpha)$ and leaves the scatter with
angle $(\vp',\beta)$. The explicit form of the matrix $D(\vp,\vp')$ is the
following
\begin{eqnarray}
D(\vp,\vp')=\hspace{10cm}\label{matriceD}\\
\mbox{\footnotesize $
\left(
\begin{array}{llll} 
\cald(\vp,\vp')&\cald(\vp,-\vp')&\cald(\vp,-\pi+\vp')&\cald(\vp,\pi-\vp')\\
\cald(-\vp,\vp')&\cald(-\vp,-\vp')&\cald(-\vp,-\pi+\vp')&\cald(-\vp,\pi-\vp')\\
\cald(-\pi+\vp,\vp')&\cald(-\pi+\vp,-\vp')&\cald(-\pi+\vp,-\pi+\vp')
&\cald(-\pi+\vp,\pi-\vp')\\
\cald(\pi-\vp,\vp')&\cald(\pi-\vp,-\vp')&\cald(\pi-\vp,-\pi+\vp')
&\cald(\pi-\vp,\pi-\vp')
\end{array}
\right)
$}
\nonumber
\end{eqnarray}
For instance (see Fig. \ref{exempletrajper})
\begin{figure}[ht]
\begin{center}
\epsfig{file=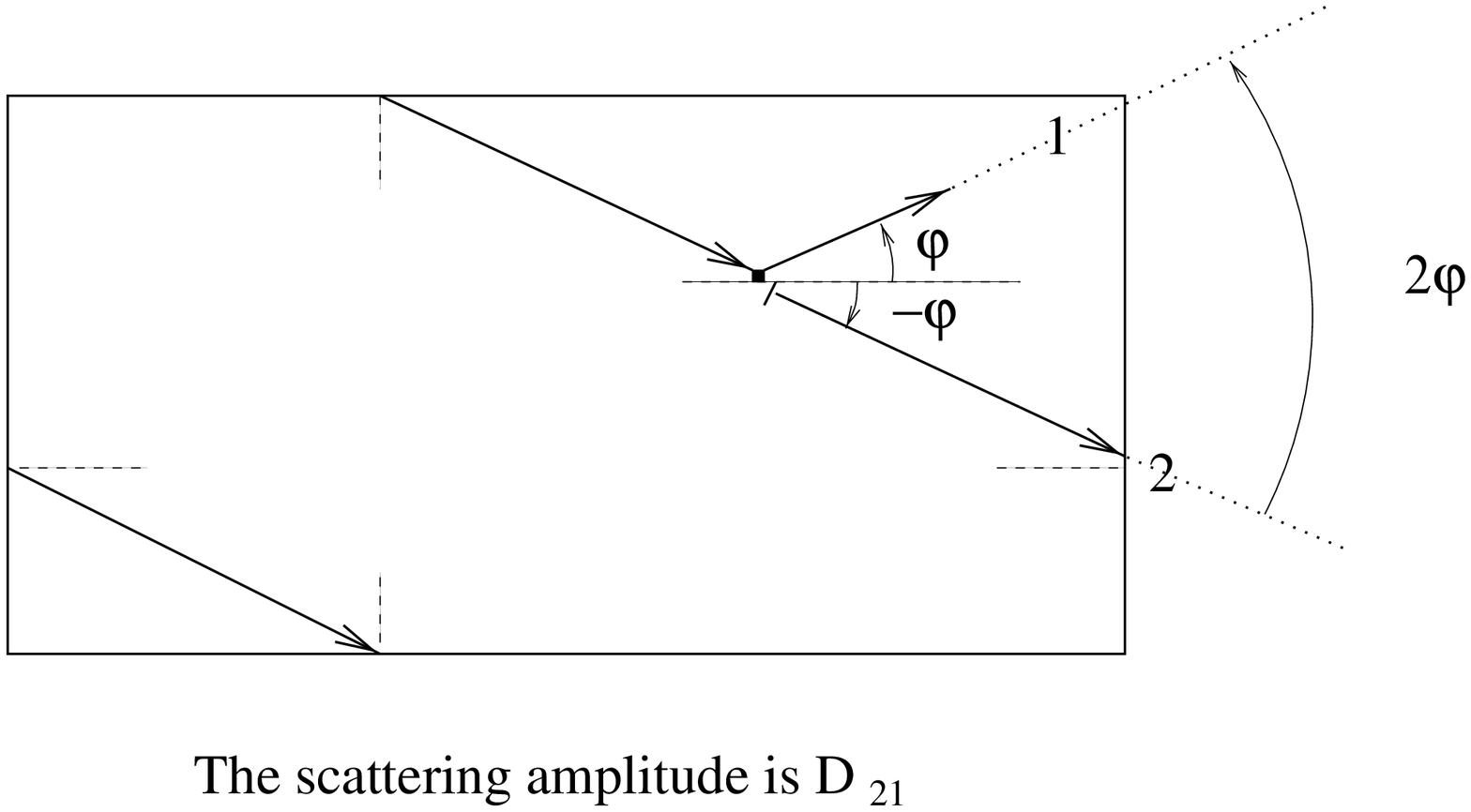,width=10cm}
\end{center}
\caption{A periodic trajectory in the rectangle with periodic boundary
  conditions}
\label{exempletrajper}
\end{figure}
if the trajectory leaves the scatter with
an angle $-\vp$ (which in our notation is angle $2$), it will come back with an
angle $\pi-\vp$, so it hits the singularity with an angle opposite to $\alpha=2$. 
If it leaves again with an angle, say $\vp$ (angle $1$),
then the scattering amplitude between the incoming and the outgoing
trajectory will be proportional to $\cald(-\vp,\vp)$, which is just
$D_{21}(\vp,\vp)$. The advantage of the matrix representation of the diffraction
coefficient is that different variables $\vp$ describe orbits with 
different lengths.
The proliferation of diffractive orbits with the same lengths is taken into
account automatically by matrix multiplication.

We now consider a trajectory made of several periodic orbits. 
Each time the trajectory leaves the scatter
with angle $\alpha$, $1\leq \alpha\leq 4$, it comes back on the scatter with the
unique angle corresponding to an additional phase $\pi$, then it leaves 
the scatter again with any angle $\beta$, $1\leq \beta\leq 4$.
In the general formula $(\ref{densitedo})$ we sum over all initial and final vectors
$\ovn_{i}, \ovn'_{j}$  with the only condition that the trajectory is
closed, which means that the outgoing angle of the first diffractive
trajectory must correspond to the incoming angle of the last diffractive
trajectory (more precisely they must differ by $\pi$).
Suppose that we consider a multiple diffractive trajectory consisting of $m$
diffractive orbits beginning and ending at the singularity and such that all
unfolded diffractive orbits are almost parallel to an angle $\vp$ in the
upper-right quadrant. Such trajectory will therefore contain a coefficient
$D_{\alpha_1 \alpha_2}(\vp,\vp)D_{\alpha_2 \alpha_3}(\vp,\vp)
\ldots D_{\alpha_{m} \alpha_1}(\vp,\vp)$.
Since we have to take into account the degeneracy of the lengths and gather
together all terms corresponding to trajectories of
same total length $l_1+\cdots+l_m$, we have to sum over
all possible incoming or outgoing angles:
\be
\sum_{\genfrac{}{}{0pt}{}{1\leq \alpha_i\leq 4}{1\leq i \leq m}}
D_{\alpha_1 \alpha_2}(\vp,\vp)D_{\alpha_2 \alpha_3}(\vp,\vp)
\ldots D_{\alpha_{m} \alpha_1}(\vp,\vp)=\tr(D^{m}(\vp,\vp)).
\end{equation}
Note that all angles are equal as we consider trajectories with equal lengths.
Of course, when $\cald$ is a constant
$\tr(D^{m}(\vp,\vp))=(4 \cald)^m$ for all integer $m\geq 1 $.

When the degeneracy has been taken into account the computation is the
same as in Section~\ref{section2} but $(g\cald)^m$ is replaced by
$\tr(D^m(\vp,\vp))$ for $m\geq 1$. 
The summation of such non-diagonal terms leads to the
multiplication of the Green function contribution $G(L)$ (given by
(\ref{GL})) by an attenuation factor similar to the one in Eq.~(\ref{sum})
\be
\label{GLper}
\left[1+\tr \left (\exp (\frac{\rho D(\vp,\vp)}{2ik}L)-1\right)\right]G(L),
\end{equation}
where as above $\rho=1/\cala$ is the density of diffraction points.

The higher-order terms in the density of states come from the contributions of
higher-order saddle-point manifolds. Let us consider a composite trajectory
consisting in $m$ diffractive orbits, among which
$m_1$ are almost parallel to the direction $\vp_1$, $m_2$ to the direction
$\vp_2$, and more generally $m_i$  are almost parallel to the direction
$\vp_i$ for $1\leq i\leq p$. Each of such trajectories is a permutation of
sequence (\ref{sequence}) but contrary to the previous Section for each
diffractive orbit there is an additional label $\alpha_i=1,2,3,4$ corresponding
to one of the 4 angles $(\ref{angles})$.

The total number of different saddle-point trajectories for $p,m$ fixed and
given $m_i$ and angle $(\vp_i, \alpha_i)$, $1\leq i\leq p$, is
\be
\label{symfac}
4^m\frac{m!}{m_1 ! \ldots m_p !},
\end{equation}
the coefficient $4$ taking into account the fact that each  orbit is
$4$ times degenerate (we will always consider $0 \leq \vp <\pd $). 

For instance at Fig.~\ref{orbpermp} there are 2 families of vectors, 
parallel to $\vp_1=0$ or its images, or parallel to $\vp_2=\pi/4$ or its 
images.
\begin{figure}[ht]
\begin{center}
\epsfig{file=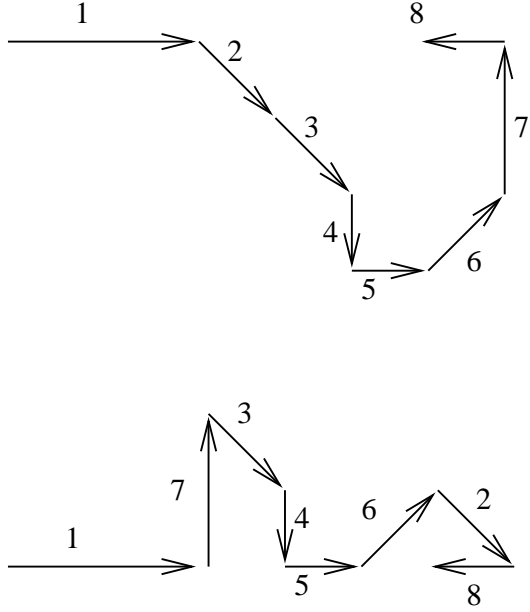,width=7cm}
\end{center}
\caption{Diffractive trajectories for $m=8, p=2, m_1=5, m_2=3$ with the same
  length.}
\label{orbpermp}
\end{figure}
At this figure two diffractive orbits are
represented: orbits $1,4,5,7,8$ are of type $1$ and orbits $2,3,6$ are of
type 2. But since now the scattering amplitude $\cald$ is not a constant,
the saddle-points will give different contributions depending on the
geometrical form of diffractive orbits. 

The $k$-th orbit is determined by a pair $(\vp_{i_k}, \alpha_k)$ with
$\alpha_k=1,\ldots, 4$ and $0\leq \vp_{i_k}<\pd $. 

For a trajectory with $p$ possible angles, $\vp_{i_k}$ can take $p$ different  
values  $\{\vp_1,\ldots ,\vp_p\}$. Each
diffraction on the scatter gives a coefficient which generic form is 
$D_{\alpha_k, \alpha_{k+1}}(\vp_{i_k}, \vp_{i_{k+1}})$, since the
coefficient $D_{ij}(\vp, \vp')$ is the scattering amplitude
corresponding to a diffraction where an orbit leaves the scatter with the
image of type $i$ of the angle $\vp$, and comes back and leaves again with
the image $j$ of the angle
$\vp'$. In order to take into account all trajectories we have to sum over all 
$\alpha_k$, $1\leq \alpha_k \leq 4$, and all $i_k$, $1\leq i_k\leq p$,  for 
$1\leq k\leq m$ (of course due to the fact that all orbits are periodic,
the index $i_{m+1}= i_1$).

Thus for fixed $m_i$ and $\vp_i$ the sum over all possible scatterings of the 
scattering amplitudes gives
\begin{equation}
R((m_1,\vp_1),\ldots, (m_p,\vp_p))=
\sum_{perm} \tr (D(\vp_{i_1},\vp_{i_2})
D(\vp_{i_2},\vp_{i_3})\ldots D(\vp_{i_m},\vp_{i_1}))
\label{R}
\end{equation}
where the summation is performed over all permutations $(i_1,...,i_m)$ 
of the sequence
(\ref{sequence}).

E.g. for the considered example $m=4$ and $m_1=m_2=2$ (see
Fig.~\ref{degeneracies}b and (\ref{example})) 
\begin{eqnarray}
&&R((2,\vp_1), (2,\vp_2))=\tr( D(\vp_{1}, \vp_{1})D(\vp_{1},\vp_{2})
D(\vp_{2}, \vp_{2})D(\vp_{2}, \vp_{1}))
\nonumber\\
&&+\tr( D(\vp_{1}, \vp_{2})D(\vp_{2},\vp_{1})
D(\vp_{1}, \vp_{2})D(\vp_{2}, \vp_{1}))
\nonumber\\
&&+\tr( D(\vp_{1}, \vp_{2})D(\vp_{2},\vp_{2})
D(\vp_{2}, \vp_{1})D(\vp_{1}, \vp_{1}))
\\
&&+\tr( D(\vp_{2}, \vp_{2})D(\vp_{2},\vp_{1})
D(\vp_{1}, \vp_{1})D(\vp_{1}, \vp_{2}))
\nonumber\\
&&+\tr( D(\vp_{2}, \vp_{1})D(\vp_{1},\vp_{2})
D(\vp_{2}, \vp_{1})D(\vp_{1}, \vp_{2}))
\nonumber\\
&&+\tr( D(\vp_{2}, \vp_{1})D(\vp_{1},\vp_{1})
D(\vp_{1}, \vp_{2})D(\vp_{2}, \vp_{2})) .
\nonumber
\end{eqnarray}
The quantity $R((m_1,\vp_1),\ldots, (m_p,\vp_p))$
replaces the corresponding quantity $(\ref{rmp})$ in Eq.~(\ref{ddomp}) 
and reduces to it if $\cald$ is a constant. 

The rest of the calculations is exactly the same as above and 
finally one gets the expression (\ref{dmain}) where
\begin{equation}
d_1(\vl)=\frac{i\cala}{2\pi}
\left[1+\tr \left (\exp (\frac{D(\vp,\vp)}{2ik\cala}L)-1\right)\right]
\end{equation}
and $d_p(\vl_1,\ldots,\vl_p)$ is the same as $(\ref{ddomp})$ with 
$R(m_1,...,m_p)$ replaced by $R$
given by $(\ref{R})$.

Using Eq.~(\ref{kpourp}) and replacing $(\ref{rmp})$ by $(\ref{R})$
 we obtain that the two-point correlation form factor
for the considered case is given by the following formulas 
\begin{eqnarray}
K(\tau)&&=<\left|1+\tr \left (\exp (-i\frac{D(\vp,\vp)}{8}\tau)-1\right )
\right|^2>
\nonumber\\
&&+\sum_{p=2}^{\infty}\frac{1}{p!}\sum_{m,n\geq p}A_{mnp}\tau^{m+n-p+1}
\label{ktrace}
\end{eqnarray}
with
\begin{eqnarray}
A_{mnp}=\frac{(-i/8)^{m}(i/8)^{n}}{m n (m+n-p-1)!}
\sum_{\genfrac{}{}{0pt}{}{\sum m_i=m}{m_i\geq 1}}
\sum_{\genfrac{}{}{0pt}{}{\sum n_j=n}{n_j\geq 1}}
\prod_{i=1}^{p}\left[\frac{(m_i+n_i-2)!}{(m_i-1)!(n_i-1)!}\right]\hspace{1cm}
\label{aperiod}\\
\times <R((m_1,\vp_{1}),\ldots, (m_p,\vp_{p}))
\bar{R}((n_1,\vp_{1}),\ldots, (n_p,\vp_{p}))>\nonumber,
\end{eqnarray}
where $R((m_1,\vp_1),\ldots, (m_p,\vp_p))$ is defined in Eq.~(\ref{R}).
The sum is taken over all partitions of $m$ and $n$ into sums of integers 
$m_i$ and $n_j$ greater than $1$ verifying 
$m_1+\cdots+m_p=m$ and $n_1+\cdots+n_p=n$.

The symbol $<\ldots >$ means the integration over all indicated angle 
variables from 0 to $\pd $
\begin{equation}
<f(\varphi_1,\ldots, \varphi_k)>=\left(\frac{2}{\pi}\right)^k\int_0^{\pd }
d\varphi_1\ldots 
\int_0^{\pd } d\varphi_k\ f(\varphi_1,\ldots, \varphi_k).
\label{fmean}
\end{equation}

\subsection{Dirichlet boundary conditions}

Dirichlet boundary conditions give rise to a slight complication as
there are now $16$ diffractive orbits associated with each couple of positive
integers $(M,N)$ (see Fig. \ref{trajdir}). 
\begin{figure}[ht]
\begin{center}
\epsfig{file=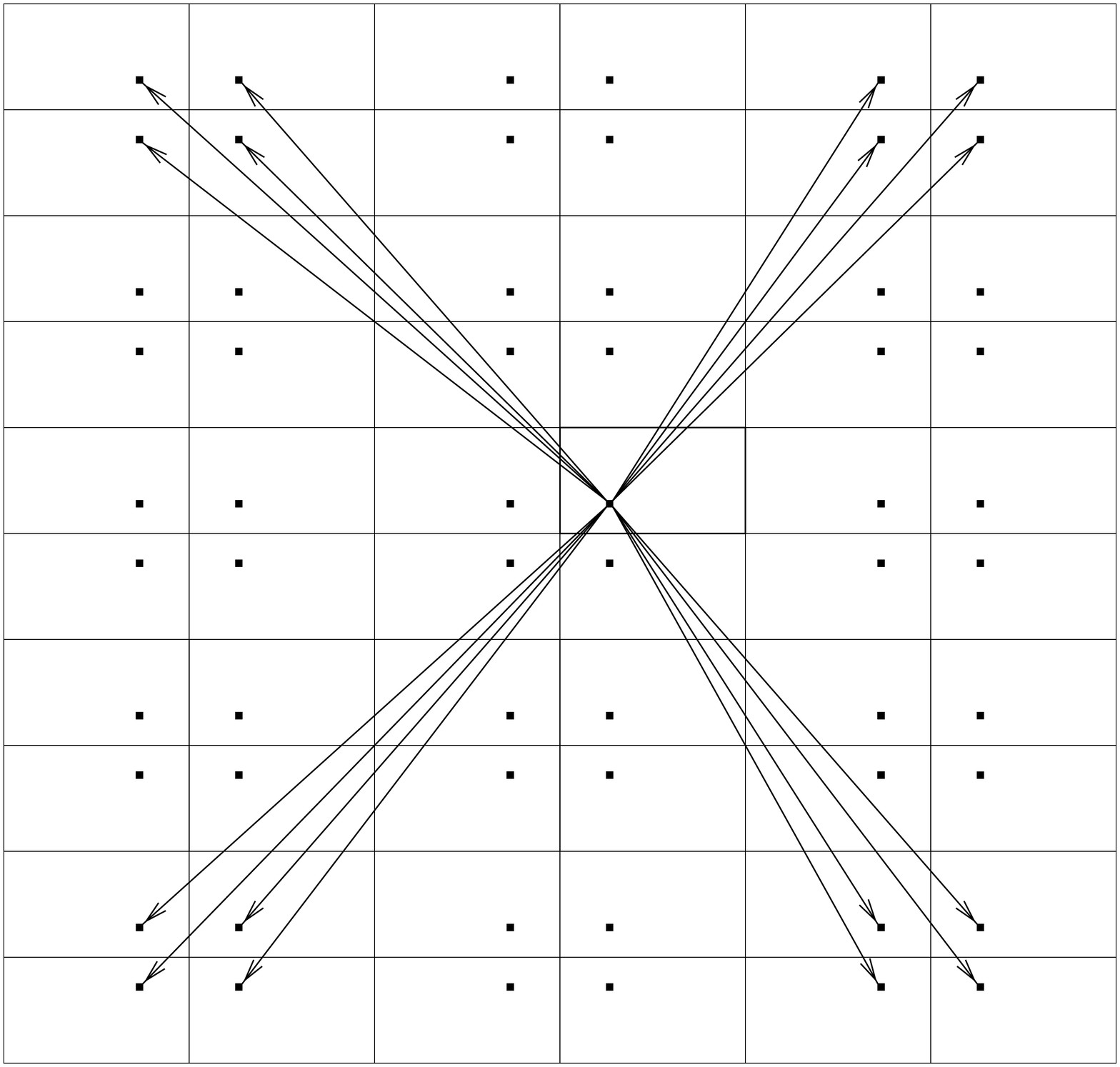,width=7cm}
\end{center}
\caption{The 16 trajectories of almost same length in the rectangle with Dirichlet
  boundary conditions.}
\label{trajdir}
\end{figure}
In Eq.~$(\ref{densitedo})$ the contribution of  a
multiple diffractive trajectory consisting in $m$ diffractive
orbits beginning and ending at the singularity and parallel to a fixed 
direction $\vp$
contains products of diffraction coefficients $\cald(\vp_i,\vp_j)$  and terms 
$\exp(i k l_i)$ coming from the Green function where $l_i$ is the length of one 
of the 16 orbits. If we expand $l_i$ for large $M,N$  each
term $\exp(i k l_i)$ can be written according to (\ref{expansionli}) as
\be 
\eps_1\eps_2
e^{i k l_i-i\frac{\pi}{2}\nu_{i}}=
e^{i k l_p}\ e^{i(\eps_1-1) k x_0 \cos\vp+ i(\eps_2-1) k y_0\sin\vp}
\label{tgreen}
\end{equation}
where $l_p$ is the length $(\ref{lepdir})$ of the closest periodic orbit and 
$\eps_1, \eps_2=\pm 1$. As in the previous Section we
will denote respectively by 1, 2, 3, and 4 the angles  $\vp$, $-\vp$,
$-\pi+\vp$ and $\pi-\vp$ corresponding to the symmetry group of the rectangle. 

For the periodic case
(see Fig. \ref{trajper}) a trajectory which leaves the scatter with
angle $(\vp, \alpha)$ comes back to it with an angle which is necessarily
the angle opposite to the 
angle $\alpha$. Therefore for periodic boundary
conditions the free motion between scatters is fixed and the proliferation
of diffractive orbits only comes from the scattering process.

For the Dirichlet boundary conditions (see Fig. \ref{trajdir}) 
by small changing of the initial angle the trajectory $(\vp,\alpha)$ 
can come back to the 
scatter with any of the 4 angles (\ref{angles})
corresponding to $\vp$, which leads to another source of 
diffractive orbits with almost the same length. 

Let us denote by $T_{\alpha \beta}(\vp)$ the coefficient
corresponding to a diffractive orbit leaving the diffractive center with 
an angle $(\vp, \alpha)$ and coming back with angle  {\it opposite to} 
$(\vp, \beta)$. Each $T_{\alpha \beta}(\vp)$ is a product of two factors
of the type
\be
\eps_1\eps_2
\exp\left[i(\eps_1-1) \phi_1+i(\eps_2-1)\phi_2\right]
\end{equation}
where angles $\phi_i=\phi_i(\vp)$ were defined in (\ref{phi1}) and (\ref{phi2}),
and $\eps_i=\pm 1$. The matrix $T=T(\vp)$
describes the free motion modification of the Green function and simple 
calculations show that it has the following form
\be
\label{matriceT}
T(\vp)=\left(
\begin{array}{cccc}
1&\bar{b}&\bar{a} \bar{b}&\bar{a}\cr
b&1&\bar{a}&b\bar{a}\cr
a b&a&1&b\cr
a&a\bar{b}&\bar{b}&1
\end{array}
\right)
\end{equation}
where $a=-e^{2i\phi_1}$ and $b=-e^{2i\phi_2}$. 

It can be checked that
\be
T_{\alpha \beta}(\vp)=V_\alpha(\vp)\bar{V}_\beta(\vp)
\label{defT}
\end{equation}
where $V_i(\phi_1(\vp),\phi_2(\vp))$ is the vector defined by
\be
\label{defV}
V(\vec{\phi})=\left(
\begin{array}{c}
\exp(-i\phi_1-i \phi_2)\cr
-\exp(-i\phi_1+i \phi_2)\cr
\exp(i \phi_1+i \phi_2)\cr
-\exp(i \phi_1-i \phi_2)
\end{array}
\right)
\end{equation}
Any diffractive trajectory can be specified by fixing initial and final
angles for each scattering. If the trajectory leaves the scatter with
an angle $(\vp, k)$ and comes back to it with angle $(\vp, l+\pi)$ (that is,
an angle opposite to $l$) it is attributed
a coefficient $T_{k l}(\vp)$ (\ref{matriceT}). If then it scatters 
to an angle $(\vp', j)$ it
gets the coefficient $D_{l j}(\vp,\vp')$ defined in (\ref{matriceD}) and so on 
(see Fig. \ref{tdtdtd}). 
\begin{figure}[ht]
\begin{center}
\epsfig{file=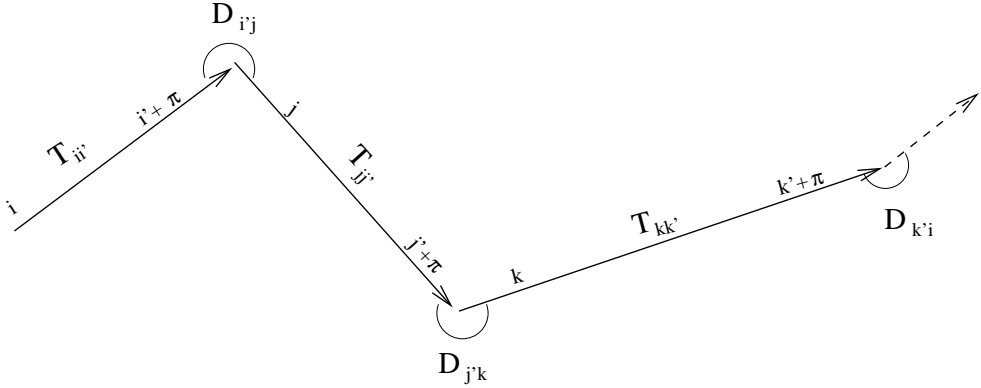,width=13cm}
\end{center}
\caption{An example of periodic trajectory for $m=3$}
\label{tdtdtd}
\end{figure}
For example, a diffractive trajectory consisting in
$m$ diffractive orbits parallel to the direction $\vp$ will be associated to
a  pre-factor
\be
\sum T_{\alpha_1 \alpha'_1} D_{\alpha'_1 \alpha_2} T_{\alpha_2 \alpha'_2}
\ldots T_{\alpha_{m} \alpha'_{m}}
D_{\alpha'_{m} \alpha_1}=\tr (T D)^{m}.
\end{equation}

As in Section~\ref{section3} assuming that the ratio $(x_0a)/(y_0b)$ is an
irrational number then as $k\rightarrow \infty$ angles
$\phi_1$ and $\phi_2$ act as independent random variables
uniformly distributed between $0$ and $\pd$. 

To take into account higher-order contributions we have to sum over 
all possible types of
vectors ($\alpha_i, \alpha'_j=1$ to $4$) and all possible angles
$\vp_{i_k}\in\{\vp_1,\ldots ,\vp_p\}$ in terms like
\begin{eqnarray}
\sum_{\alpha_i, \alpha'_j}D_{\alpha_1, \alpha'_1}(\vp_{i_1},\vp_{i_2}) T_{\alpha'_1, \alpha_2}(\vp_{i_2})
D_{\alpha_2, \alpha'_2}(\vp_{i_2},\vp_{i_3}) T_{\alpha'_2, \alpha_3}(\vp_{i_3})
\ldots \nonumber\\
=\tr (D(\vp_{i_1},\vp_{i_2})T(\vp_{i_2})D(\vp_{i_2},\vp_{i_3})
T(\vp_{i_3})\ldots )
\end{eqnarray}
under the constraint that each angle $\vp_{i}$ appears exactly $m_i$ times.
Using the property (\ref{defT}) one can rewrite the above expression by
introducing the effective diffraction coefficient $S$ defined by
\be
\label{S}
S(\vp, \vp')={}^{t}\bar{V}_\alpha(\varphi) 
D_{\alpha \beta}(\varphi,\varphi')V_\beta(\varphi')
\end{equation}
which describes all possibilities of scattering and free motion between 
the scatter. Since $S(\vp, \vp')$ is a number we get
\begin{eqnarray}
&&\tr (D(\vp_{i_1},\vp_{i_2})T(\vp_{i_2})D(\vp_{i_2},\vp_{i_3})
T(\vp_{i_3})\ldots )\nonumber\\
&&\hspace{3cm}=S(\vp_{i_1},\vp_{i_2})
S(\vp_{i_2},\vp_{i_3})\ldots S(\vp_{i_m},\vp_{i_1})
\end{eqnarray}
Let us define a quantity similar to Eq.~(\ref{R})
\be
R((m_1,\vp_1),\ldots, (m_p,\vp_p))=\sum_{perm}S(\vp_{i_1},\vp_{i_2})
S(\vp_{i_2},\vp_{i_3})\ldots S(\vp_{i_m},\vp_{i_1}),
\label{Rtilde}
\end{equation}
where the sum is taken over all permutations $(i_1, ..., i_m)$
 of the sequence $(\ref{sequence})$
and it is implicitly assumed  the set $\vp_i$ contains $m_1$ terms of 
$\vp_1$, $m_2$ terms of $\vp_2$,$\ldots$, $m_p$ terms of $\vp_p$. 
It can be checked that Eq. $(\ref{Rtilde})$ reduces to 
$(\ref{Rdircst})$ if $\cald$ is a constant, since we have
\be
\sum_{i,j}{}^{t}\bar{V}_{i}(\varphi) V_{j}(\varphi)= g r(\vp)
\end{equation}
with $g=4$ and $r(\vp)$ given by $(\ref{rvarphi})$.
The  final expression for the two-point correlation form factor, $K(\tau)$,
follows from exactly the same consideration as in  previous Sections.
It reads
\begin{eqnarray}
K(\tau)=<<\left|\exp\left(-i\frac{S(\vp,\vp)}{8}\tau\right)\right|^2>>\nonumber \\
+\sum_{p=2}^{\infty}\frac{1}{p!}\sum_{m,n\geq p}A_{mnp}\tau^{m+n-p+1}
\label{kdirichlet}
\end{eqnarray}
where
\begin{eqnarray}
A_{mnp}=\frac{(-i/8)^{m}(i/8)^{n}}{m n (m+n-p-1)!}
  \sum_{\genfrac{}{}{0pt}{}{\sum m_i=m}{m_i\geq 1}}
\sum_{\genfrac{}{}{0pt}{}{\sum n_j=n}{n_j\geq 1}}
\prod_{i=1}^{p}\left[\frac{(m_i+n_i-2)!}{(m_i-1)!(n_i-1)!}\right]
\nonumber\\
\times <<R((m_1,\vp_{1}),\ldots, (m_p,\vp_{p}))
\bar{R}((n_1,\vp_{1}),\ldots,(n_p,\vp_{p})>>.
\label{adirichlet}
\end{eqnarray}
As usual the sum is taken over all partitions of $m$ and $n$ into sums of
$p$ positive integers. The symbol $<<\ldots>>$ denotes the
double average over random phases $\phi_i$ as in Eq.~(\ref{phimean}) and
geometrical angles $\varphi_j$ as in Eq.~(\ref{fmean}).

\section{Conclusion}
\label{section5}

We have discussed the spectral statistics of a rectangular billiard with 
a small-size
impurity inside and developed the method which permits the explicit construction
of perturbation expansion of the two-point correlation form factor for this
system, $K(\tau)$, into series of $\tau$
\begin{equation}
K(\tau)=\sum_{n}c_n\tau^n.
\end{equation}
Using the method of Ref.~\cite{bogomolny2} we demonstrate that after the
summation over diffractive orbits with approximately the same lengths the
oscillating part of the density of states can be written in the form
(\ref{dmain})
\be
d^{(osc)}(E)=\sum_{p=1}^{\infty}
\sum_{L_1<\ldots<L_p}d_{p}(\vec{L}_1,\ldots,\vec{L}_p)G(L_1)\ldots G(L_p)+c.c.,
\label{dmain2}
\end{equation}
with
\begin{eqnarray}
&&d_p(\vec{L}_1,\ldots,\vec{L}_p)=\frac{L_1+\ldots+L_p}{4\pi k} 
\sum_{m=p}^{\infty}\frac{1}{m}\nonumber\\
&&\times \sum_{\genfrac{}{}{0pt}{}{m_1+\ldots +m_p=m}{m_i\geq 1}}
R((m_1,\vp_1),\ldots,(m_p,\vp_p))\nonumber\\
&&\times (\frac{L_1}{2ik\cala})^{m_1-1}\frac{1}{(m_1-1)!}\ldots
(\frac{L_p}{2ik\cala})^{m_p-1}\frac{1}{(m_p-1)!},
\label{ddomp2}
\end{eqnarray}
where the summation is taken over all partitions of $m$ into sums of $p$
positive integers $m=m_1+\ldots+m_p$ 
and $R((m_1,\vp_1),\ldots,(m_p,\vp_p))$ $(\ref{Rtilde})$
is the contribution of all diagrams
describing the scattering process of  composite diffractive orbits
consisting of  $m_1$ orbits of type 1, $m_2$ orbits of type 2, and so on. 

To construct such diagrams it is convenient first to write down all
permutations of the sequence (\ref{sequence}) of $m$ elements with $m_i$ elements 
of type $i$. The total number of these permutations is
\be
N(m_1,\ldots m_p)=\frac{m!}{m_1!\ldots m_p!}.
\end{equation}
Each symbol $i$ represents the angle $\vp_i$ and the constructed sequence of
symbols  is considered as the representation of the
scattering process. The symmetry group of the rectangular billiard 
leads to the necessity to consider together all symmetry partners of a given
trajectory, that is all possible choices for each diffractive orbit.

For periodic boundary conditions there are 4 different
possibilities (\ref{angles}) as indicated at Fig.~\ref{trajper}, whereas
for Dirichlet conditions it gives rise to  16 trajectories with
approximately the same lengths as at Fig.~\ref{trajdir}. To take into
account all these degenerate and almost degenerate trajectories it is useful
to use the matrix formalism developed in the previous Sections. All the 
scattering processes between 2 consecutive angles $\vp$ and $\vp'$ are
described by the matrix $D(\vp, \vp')$ defined in Eq.~(\ref{matriceD}) for
periodic boundary conditions, and by the quantity $S(\vp, \vp')$ 
defined in Eq.~(\ref{S}) for Dirichlet ones.

The quantity $R((m_1,\vp_1),\ldots,(m_p,\vp_p))$ is the sum of 
such contributions for all
permutations of our standard sequence (\ref{sequence}), that is all possible
orderings of the sequence of diffractive orbits.

The knowledge of  $R((m_1,\vp_1),\ldots,(m_p,\vp_p))$ permits easily the 
computation of the two-point correlation form factor in all cases. 
The advantage of the representation
(\ref{dmain2}) is that different  terms in it are non-coherent and the two-point  
correlation function is simply given by the sum of the squares of each terms as
in  Eq.~(\ref{r2final}). This
`generalized diagonal approximation' leads to explicit formulas for the
form factor Eqs.~(\ref{ktrace}), (\ref{aperiod}) for the periodic boundary
conditions and Eqs.~(\ref{kdirichlet}) and (\ref{adirichlet}) for the
Dirichlet ones which solve the problem of the summation over non-diagonal terms
for integrable systems.

\section*{Appendix}

The exact expression for the two-point correlation function, $R_2(\eps)$,
for a rectangular billiard perturbed by a small-size impurity with constant
diffraction coefficient has been derived in Ref.~\cite{gerland}.
The aim of this Appendix is to obtain its perturbation expansion for large
$\eps$ and the corresponding expansion of the two-point correlation
form factor, $K(\tau)$, into series of $\tau$. 

According to Eq. (143) in \cite{gerland} (with slight changing of the
notations) the two-point correlation function can be written as multiple integrals
\begin{eqnarray}
&&R_2(\eps)-1=\int_{0}^{\infty}d\alpha_1 \int_{0}^{\infty} d\alpha_2\ 
  e^{-\pi \eps <J(r\alpha_1, r\alpha_2)>}\\
&&\times [<rJ_0(2r\sqrt{\alpha_1\alpha_2})e^{ir(\alpha_1+\alpha_2)}>^2
+ <rJ_1(2r\sqrt{\alpha_1\alpha_2})e^{ir(\alpha_1+\alpha_2)}>^2 ] 
+\cc \nonumber 
\end{eqnarray}
where 
\begin{eqnarray}
&&J(\alpha_1, \alpha_2)=(\alpha_1-\alpha_2)(1+\frac{i}{\pi v'}
+2i 
e^{i\alpha_2}\int_{\alpha_1}^{\infty}J_0(2 \sqrt{\alpha_2 t})e^{it}dt)
\nonumber\\
&-&i[\alpha_1 J_0(2\sqrt{\alpha_1\alpha_2})+\sqrt{\alpha_1\alpha_2}
J_1(2\sqrt{\alpha_1\alpha_2})]e^{i(\alpha_1+\alpha_2)}
\label{defn}
\end{eqnarray}
Here $J_0(x)$ and $J_1(x)$ are the usual Bessel functions. The renormalized
coupling constant $v'$ is connected with the diffraction coefficient, $\cald$,
by the relation
\be
\label{coeffd}
\cald=\frac{1}{\frac{i}{4}+\frac{1}{4\pi v'}}.
\end{equation}
The variable $r=r(\phi_1,\phi_2)$ depends on the boundary conditions 
(see \cite{gerland})
\be
r(\phi_1,\phi_2)=\left \{ \begin{array}{cl}
1&\mbox{for periodic boundary conditions}\\
4\sin^2(\phi_1)\sin^2(\phi_2)&\mbox{for Dirichlet boundary conditions}\\
4\cos^2(\phi_1)\cos^2(\phi_2)&\mbox{for Neumann boundary conditions}
\end{array} \right . ,
\end{equation}
where $\phi_1$ and $\phi_2$ are independent angles distributed uniformly
between $0$ and $\pi/2$. The symbol $<\ldots>$ denotes the mean values over
these angles
\be
<f(\phi_1,\phi_2>=\frac{4}{\pi^2}\int_{0}^{\pi/2}d\phi_1
\int_{0}^{\pi/2}d\phi_2 f(\phi_1,\phi_2).
\end{equation}
In particular we have $<r(\phi, \phi)>=1$.
Changing variables $\eps$ to $x$ and $\alpha_1, \alpha_2$ to $\alpha_1/x, \alpha_2/x$, 
one gets
\begin{eqnarray}
\label{rmun}  
&&R_2(x)-1=\int_{0}^{\infty}d\alpha_1\int_{0}^{\infty} d\alpha_2
e^{-\pi x <J(\frac{r\alpha_1}{x},\frac{r\alpha_2}{x})>}\\
&&\times [<rJ_0(\frac{2r\sqrt{\alpha_1\alpha_2}}{x})
e^{ir\frac{\alpha_1+\alpha_2}{x}}>^2
+<rJ_1(\frac{2r\sqrt{\alpha_1\alpha_2}}{x})
e^{ir\frac{\alpha_1+\alpha_2}{x}}>^2]+\cc.
\nonumber
\end{eqnarray}
To compute the expansion of the above expressions into powers of $1/x$ it is
convenient to use the transformations proposed in \cite{Berko}.

From $(\ref{defn})$ it follows that
\be
\label{partialun}
\left(\frac{\partial}{\partial\alpha_1}
  +\frac{\partial}{\partial\alpha_2}\right)
  x J(\frac{\alpha_1}{x},\frac{\alpha_2}{x})=
  2J_0(\frac{2\sqrt{\alpha_1\alpha_2}}{x})e^{i\frac{\alpha_1+\alpha_2}{x}}
\end{equation}
and
\be
\label{partialdeux}
\frac{\partial}{\partial x}
  x J(\frac{\alpha_1}{x},\frac{\alpha_2}{x})=
-2i  \frac{\sqrt{\alpha_1\alpha_2}}{x}
J_1(\frac{2\sqrt{\alpha_1\alpha_2}}{x})e^{i\frac{\alpha_1+\alpha_2}{x}}.
\end{equation}
Using the fact that $J'_0(x)=-J_1(x)$ and $J'_1(x)=J_0(x)-J_1(x)/x$,
one can check that
\be
\left(\frac{\partial}{\partial\alpha_1}
  +\frac{\partial}{\partial\alpha_2}\right)^2
 x J(\frac{\alpha_1}{x},\frac{\alpha_2}{x})=\frac{x^2}{\alpha_1\alpha_2}
 \frac{\partial^2}{\partial x^2}
  x J(\frac{\alpha_1}{x},\frac{\alpha_2}{x}).
\end{equation}  
Therefore, 
\begin{eqnarray}
\left[\left(\frac{\partial}{\partial\alpha_1}
  +\frac{\partial}{\partial\alpha_2}\right)^2-\frac{x^2}{\alpha_1\alpha_2}
 \frac{\partial^2}{\partial x^2}\right]
 e^{-\pi x <J(\frac{r\alpha_1}{x},\frac{r\alpha_2}{x})>}=\hspace{3cm}\nonumber\\
\pi^2\left[\left( \left(\frac{\partial}{\partial\alpha_1}
  +\frac{\partial}{\partial\alpha_2}\right)
  x <J(\frac{r\alpha_1}{x},\frac{r\alpha_2}{x})>\right)^2\right.
\hspace{1cm}  \\
 - \left.\frac{x^2}{\alpha_1\alpha_2}
 \left(\frac{\partial}{\partial x}
x <J(\frac{r\alpha_1}{x},\frac{r\alpha_2}{x})>\right)^2\right]
e^{-\pi x <J(\frac{r\alpha_1}{x},\frac{r\alpha_2}{x})>}\nonumber
\end{eqnarray}
and according to Eqs. $(\ref{partialun})$ and $(\ref{partialdeux})$ the
previous equation can be transform as follows
\begin{eqnarray}
&& \frac{1}{4\pi^2}\left[\left(\frac{\partial}{\partial\alpha_1}
  +\frac{\partial}{\partial\alpha_2}\right)^2-\frac{x^2}{\alpha_1\alpha_2}
 \frac{\partial^2}{\partial x^2}\right]
 e^{-\pi x <J(\frac{r\alpha_1}{x},\frac{r\alpha_2}{x})>}=\\
&& \left [ <rJ_0(\zeta)
e^{ir\frac{\alpha_1+\alpha_2}{x}}>^2
+<rJ_1(\zeta)
e^{ir\frac{\alpha_1+\alpha_2}{x}}>^2 \right ]  
e^{-\pi x <J(\frac{r\alpha_1}{x},\frac{r\alpha_2}{x})>},
\nonumber
\end{eqnarray}
where $\zeta=2r\sqrt{\alpha_1\alpha_2}/x$.

The right-hand side of this expression is the pre-factor which appears in 
Eq.~$(\ref{rmun})$ and, consequently,  one can express the two-point
correlation function in the form
\begin{eqnarray}
R_2(x)-1&=&\int\limits_{0}^{\infty}\frac{d\alpha_1 d\alpha_2}{4\pi^2 x^2}
\left[\left(\frac{\partial}{\partial\alpha_1}
  +\frac{\partial}{\partial\alpha_2}\right)^2-\frac{x^2}{\alpha_1\alpha_2}
 \frac{\partial^2}{\partial x^2}\right]
e^{-\pi x <J(\frac{r\alpha_1}{x},\frac{r\alpha_2}{x})>}
\nonumber\\
&+&\cc
\label{deuxparties}
\end{eqnarray}
The  derivatives with respect to $\alpha_1$ and $\alpha_2$ can be 
computed by integration by parts. It gives
\begin{eqnarray}
\int_{0}^{\infty}d\alpha_1\ 
\frac{\partial}{\partial\alpha_1}
\left(\frac{\partial}{\partial\alpha_1}
+\frac{\partial}{\partial\alpha_2}\right)
e^{-\pi x <J(\frac{r\alpha_1}{x},\frac{r\alpha_2}{x})>}=\hspace{3cm}\nonumber\\
2\pi\left<r e^{i r \alpha_2/x}\right>
e^{-\pi\alpha_2
\left(1-\frac{i}{\pi v'}\right)}
\end{eqnarray}
and 
\begin{eqnarray}
\int_{0}^{\infty}d\alpha_2\ 
\frac{\partial}{\partial\alpha_2}
\left(\frac{\partial}{\partial\alpha_1}
+\frac{\partial}{\partial\alpha_2}\right)
e^{-\pi x <J(\frac{r\alpha_1}{x},\frac{r\alpha_2}{x})>}=\hspace{3cm}\nonumber\\
2\pi\left<r e^{i r \alpha_1/x}\right>
e^{-\pi\alpha_1
\left(1+\frac{i}{\pi v'}\right)}.
\end{eqnarray}
Therefore
\begin{eqnarray}
&&R_0(x)=\int_{0}^{\infty}\frac{d\alpha_1 d\alpha_2}{4\pi^2 x^2}
\left(\frac{\partial}{\partial\alpha_1}
  +\frac{\partial}{\partial\alpha_2}\right)^2
e^{-\pi x <J(\frac{r\alpha_1}{x},\frac{r\alpha_2}{x})>}
+\cc=\nonumber\\
&&\hspace{3cm}\frac{1}{\pi x}\left<
\frac{ir\left(r+i\pi x\right)}{\left(r+i\pi x\right)^2-
\left(\frac{x}{v'}\right)^2}\right>+\cc
\label{partieun}
\end{eqnarray}
We are interested in the Fourier transform
\be
\label{fourierk}
K(\tau)=\int_{-\infty}^{\infty}d x\ (R_2(x)-1) e^{2i\pi\tau x}
\end{equation}
of the two-point correlation function. 

The term (\ref{partieun}) gives the following contribution
\be
\label{kpmun}
K_0(\tau)=<e^{2i\pi\tau x_{+}}+e^{2i\pi\tau x_{-}}>-2.
\end{equation}
where
\be
\label{poles}
x_{\pm}=\frac{r}{\pm\frac{1}{v'}-i\pi}
\end{equation}
are the poles in $(\ref{partieun})$.
Using the definition (\ref{coeffd}) of $\cald$,  Eqs.~$(\ref{kpmun})$
and $(\ref{poles})$ lead to 
\be
\label{kpm}
K_0(\tau)=<e^{-\frac{i r \cald g}{8}\tau}
+e^{\frac{i r \bar{\cald} g}{8}\tau}>-2,
\end{equation}
where we introduce $g=4$ to be consistent with notations in Section~\ref{section2}.

The other contributions come from the second derivatives with respect to
$x$ in $(\ref{deuxparties})$. 
Using expansion (112) of  Ref.~\cite{gerland} one gets
\begin{eqnarray}
  x <J(\frac{r\alpha_1}{x},\frac{r\alpha_2}{x})>&=&
\alpha_1+\alpha_2  +\frac{i}{\pi v'}(\alpha_1-\alpha_2)\\
&-&2i\sum_{m,n=0}^{\infty}
\frac{(i\alpha_1)^{m+1}(i\alpha_2)^{n+1}}{x^{m+n+1}}
\frac{(m+n)!<r^{m+n+2}>}{m!(m+1)!n!(n+1)!}.\nonumber
\end{eqnarray}
The expansion of the exponential of this quantity yields
\begin{eqnarray}
e^{-\pi x <J(\frac{r\alpha_1}{x},\frac{r\alpha_2}{x})>}=
e^{-\pi(\alpha_1+\alpha_2)}e^{-\frac{i}{v'}(\alpha_1-\alpha_2)}
\sum_{p=0}^{\infty}\frac{(2i\pi)^p}{p!}\sum_{m,n\geq 0}
\hspace{2cm}\nonumber\\
\sum_{\genfrac{}{}{0pt}{}{m_1+\cdots+m_p=m}{n_1+\cdots+n_p=n}}
\prod_{i=1}^{p}\frac{C_{m_i+n_i}^{m_i}}{(m_i+1)!(n_i+1)!}<r^{m_i+n_i+2}>
\frac{(i\alpha_1)^{m+p}(i\alpha_2)^{n+p}}{x^{m+n+p}}.
\end{eqnarray}
As 
\be
\int_{0}^{\infty}t^{\alpha}e^{-\sigma t}d t=\frac{\alpha!}{\sigma^{\alpha+1}},
\end{equation}
the integration over $\alpha_i$ gives the expansion of the two-point
correlation function into series of $1/x$
\be
R_2(x)-1=\sum_{p=0}^{\infty} R_p(x),
\end{equation}
where $R_0(x)$ is given by (\ref{partieun}) and 
\be
\label{sumpun}
R_p(x)=-\frac{(2i\pi)^p}{p!} 
\sum_{m,n\geq 0}\frac{(m+n+p+1)!A_{mnp}}
{4\pi^2 x^{m+n+p+2}}
\left(-\frac{\cald g}{16\pi}\right)^{m+p}
  \left(\frac{\bar{\cald}g}{16\pi}\right)^{n+p}
\end{equation}
where 
\be
\label{amnpapp}
A_{mnp}= \frac{(m+p-1)!(n+p-1)!}{(m+n+p-1)!}
\sum_{m_i, n_j\geq 0}
\prod_{i=1}^{p}\left[\frac{C_{m_i+n_i}^{m_i}<r^{m_i+n_i+2}>}
{(m_i+1)!(n_i+1)!}\right].
\end{equation}
The expression $(\ref{amnpapp})$ for $A_{mnp}$ is the same as in 
Eq.~(\ref{amndir}).

Taking the Fourier transform $(\ref{fourierk})$ of $(\ref{sumpun})$, 
we get the corresponding expansion of the two-point correlation form factor.
The term corresponding to $p=1$ can be transformed to the form
\be
\label{kun}
K_1(\tau)=<\left(e^{-\frac{i r \cald g}{8}\tau}-1\right)
\left(e^{\frac{i r \bar{\cald} g}{8}\tau}-1\right)>.
\end{equation}
The sum of the contributions  (\ref{kpm}) and
(\ref{kun}) (plus a $1$ coming from the $\delta$-function in $R_2(x)$) gives
\be
<e^{-\frac{i r \cald g}{8}\tau}e^{\frac{i r \bar{\cald} g}{8}\tau}>=
<e^{-\frac{|\cald|^2 g r}{16}\tau}>.
\end{equation}
The Fourier transform of the sum (\ref{sumpun}) gives the
terms for $p\geq 2$ 
\be
\sum_{p=2}^{\infty}\frac{1}{p!}\sum_{m,n\geq 0}A_{mnp}
\left(-\frac{i \cald g}{8}\right)^{m+p}
  \left(\frac{i\bar{\cald}g}{8}\right)^{n+p}
\tau^{m+n+p+1} ,   
\end{equation}
which coincides with the result (\ref{ffdir}).

\end{document}